\documentclass[conference]{IEEEtran}
\usepackage{bm}
\usepackage{graphicx}
\usepackage{color}
\usepackage{amssymb}

\makeatletter
\def\ps@headings{%
\def\@oddhead{\mbox{}\scriptsize\rightmark \hfil \thepage}%
\def\@evenhead{\scriptsize\thepage \hfil \leftmark\mbox{}}%
\def\@oddfoot{}%
\def\@evenfoot{}}
\makeatother
\newtheorem{defi}{Definition}
\newtheorem{prop}{Proposition}
\newtheorem{coro}{Corollary}

\begin{document}
\title{GeoQuorum: Load Balancing and Energy Efficient Data Access in Wireless Sensor Networks}

\author{\IEEEauthorblockN{Jun Luo ~~~~~~~~ Ying He}
\IEEEauthorblockA{School of Computer Engineering \\
                  Nanyang Technological University (NTU), Singapore \\
Email: \{junluo,yhe\}@ntu.edu.sg
}
}
\maketitle

\begin{abstract}
  When data productions and consumptions are heavily unbalanced and when the origins of data queries are spatially and temporally distributed, the so called \textit{in-network data storage paradigm} supersedes the conventional \textit{data collection paradigm} in \textit{wireless sensor networks} (WSNs). In this paper, we first introduce \textit{geometric quorum systems} (along with their metrics) to incarnate the idea of in-network data storage. These quorum systems are ``geometric'' because curves (rather than discrete node sets) are used to form quorums. We then propose GeoQuorum as a new quorum system, for which the quorum forming curves are parameterized. Though our proposal stems from the existing work on using curves to guide data replication and retrieval in dense WSNs, we significantly expand this design methodology, by endowing GeoQuorum with a great flexibility to fine-tune itself towards different application requirements. In particular, the tunability allows GeoQuorum to substantially improve the load balancing performance and to remain competitive in energy efficiency. Both our analysis and simulations confirm the performance enhancement brought by GeoQuorum.
\end{abstract}

\section{Introduction} \label{sec:intro}
  Since their inception, \textit{wireless sensor networks} (WSNs) bear the task of intensive data collection through their large scales and dense deployments, which represents a significant improvement over traditional sensing systems \cite{AkyildizSSC-CM02}. However, the low-cost devices (those tiny sensor nodes) involved in a WSN are also constraining factors to their missions: the limited energy storage of a node heavily confines its ability of intensively transmitting the acquired data. As a result, we have been witnessing a great volume of research developments, aiming at tackling the conflict between the need for low power operations and the requirements of large scale data gathering.

  The related research proposals have been mainly focusing on two issues: namely load balancing and energy efficiency \cite{AhnK-MobiHoc06}. Whereas the former is concerning not imposing a too heavily communication load upon a small set of nodes, the latter is managing to reduce the total communication load taken by the whole WSN. From a conventional point of view, a WSN needs to collect data from a large set of nodes to a particular (often small) set of nodes. The resulting \textit{convergecast} type of data transmission pattern under this assumption makes the above two objectives contradict each other. For example, shortest path routing, as the most energy efficient communication protocol, may lead to very unbalanced load distribution, hence ``kill'' those heavily loaded nodes long before other nodes running out of their battery \cite{ChangT-Infocom00,LuoH-ToN10}.

\begin{figure}[htb]
     \begin{center}
       \includegraphics[width=\columnwidth]{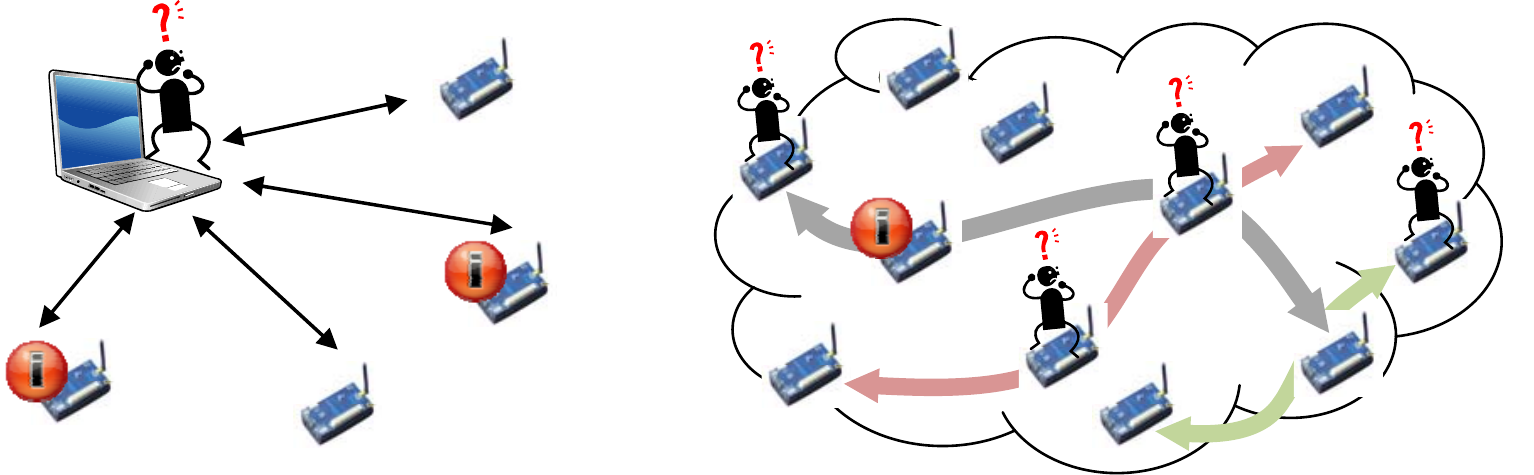}
       \parbox{\columnwidth}{\parbox{.4\columnwidth}{\center\scriptsize(a) Convergecast}
                             \parbox{.6\columnwidth}{\center\scriptsize(b) In-network data storage}}
     \caption{Comparison of two data access paradigms.}
     \label{fig:qillustrate}
    \end{center}
    \vspace{-1ex}
  \end{figure}

  Fortunately, the convergecast data collection paradigm is not the only way of acquiring data from a WSN. In particular, when data productions and consumptions are heavily unbalanced (e.g. queries from human users to the sensory data are less frequent than generation of those data), or when the queries may originate in a spatially and temporally distributed manner, other data access schemes involving \textit{in-network data storage} (e.g., \cite{RatnasamyKYYEGS-WSNA02,ShengLM-MobiHoc06}) are more meaningful. More importantly, we may strike a better tradeoff between load balancing and energy efficiency under such a data access paradigm, as our paper will demonstrate. Here we simply provide an illustration of a data access paradigm using in-network data storage in Fig.~\ref{fig:qillustrate}.
  It is clear that, whereas the convergecast collects the data at a single point, the in-network storage replicates data at various nodes, to which a later data query is directed. The latter paradigm endows a greatly flexibility to the data access: it can be performed whenever and wherever needed. Last but not not least, the in-network data storage is technically enabled by the new developments on flash memory storage \cite{MathurDGS-IPSN06}, which is ultra energy-efficient and has huge capacity.

  In this paper, we are focusing on a particular design methodology, \textit{quorum systems}, under the in-network data storage paradigm. Based on this methodology, data produced by sensor nodes and queries generated by human users are both directed to certain \textit{quorums} (subsets of nodes). As the intersection between quorums are guaranteed in the design phase, users may access the sensory data without directly communicating with the sources that generate those data. Although quorum systems exist in distributed systems \cite{MalkhiR-DC98} and have been applied to wired and wireless networking (e.g., \cite{ZhouSR-TCS02,HaasL-ToN99,LuoEH-TMC04}), we are reviving them in the sensor networking scenarios. Moreover, our design method, namely, \textit{geometric quorum systems} (GQS), leverages on the recent developments in using geometric principles to guide the protocol implementations in WSNs, e.g., \cite{SarkarZG-MobiCom06,ZengSLGG-Infocom10}. In particular, we propose GeoQuorum where the quorums are formed by parameterized curves. Tuning the parameters that determine the quorums allows us to flexibly identify desired tradeoffs between load balancing and energy efficiency. Through both analysis and simulations, we further demonstrate that our design outperforms the existing ones in terms of both load balancing and energy efficiency. In summary, our main contributions are:
  \begin{itemize}
    \item A formal definition of GQS and the related metrics.
    \item A thorough analysis of the existing quorum system designs for WSNs against the defined metrics.
    \item A general conformal geometry based quorum design methodology that applies to WSNs with any shape of the network areas.
    \item A specific quorum system, GeoQuorum, formed by parameterized curves, allowing a flexible tradeoff to be made between load balancing and energy efficiency.
  \end{itemize}

  The remaining of this paper goes as follows. In Sec.~\ref{sec:qs}, we define quorum systems (in the traditional sense) and their metrics, and we also briefly review the application of quorum systems in networked settings, in particular a recent geometry-based quorum system design. We focus on GQS in Sec.~\ref{sec:gqs}. Starting with the conformal geometry basics and network model in Sec.~\ref{sec:bkgccg}, we formally define GQS in Sec.~\ref{sec:gqsmd}, we then analyze the performance of existing designs in Sec.~\ref{sec:gqsals} and propose GeoQuorum in Sec.~\ref{sec:gqsgq}. We report the simulation results in Sec.~\ref{sec:sim} and conclude our paper in Sec.~\ref{sec:con}.

\section{Fundamental of Quorum Systems} \label{sec:qs}
  \subsection{Basic Definitions} \label{sec:qsbs}
    Quorum systems represent a fundamental abstraction for coordination among the nodes of a distributed system (e.g., a set of networked nodes). In its traditional sense, a quorum system is defined upon a finite set (also termed \textit{universe}) $\mathcal{U} = \{u_1, u_2, \cdots, u_n\}$ of nodes.  In particular, the following definition characterizes a quorum system \cite{MalkhiR-DC98}.
    \begin{defi}[Quorum System] A quorum system $\mathcal{Q} \subset 2^\mathcal{U}$ is a set of subsets of $\mathcal{U}$ such that every two subsets intersect. Each $Q \in \mathcal{Q}$ is called a \textit{quorum}.
    \end{defi}

    Given a quorum system $\mathcal{Q}$, networked nodes may make use of it to perform coordinations, such as sharing information. A node may choose to \textit{access} a quorum by either \textit{writing} to or \textit{reading} from it. Thanks to the intersection property, a read access will find the desired data from some quorum that stores the data written by another node. Note that the goal of a read access is to search for the data, whereas data delivery is carried out by a certain routing scheme that is independent of the quorum system. Taking into account the inherent asymmetry between read and write accesses, we may redefine the quorum system in a asymmetric fashion as follows \cite{LuoEH-TMC04}, the earlier definition hence specifies \textit{symmetric} quorum systems.
    \begin{defi}[Asymmetric Quorum System] An asymmetric quorum system $\mathcal{Q} \subset 2^\mathcal{U}$  consists of two disjoint sets, $\mathcal{Q}^R$ and $\mathcal{Q}^W$, of subsets of $\mathcal{U}$, such that each subset in $\mathcal{Q}^R$ intersects every subset in $\mathcal{Q}^W$. Each subset in $\mathcal{Q}^R$ (resp. $\mathcal{Q}^W$) is called a \textit{read} (resp. \textit{write}) \textit{quorum}.
    \end{defi}

  \subsection{Metrics on Quorum Systems}
    We introduce two metrics to measure the performance of quorum systems, namely, \textit{load} and \textit{robustness}.

    \subsubsection{Load} This metric measures the computational load taken by individual nodes due to their participation in various quorums. Obviously, it depends not only on how a quorum system is constructed, but also on what strategy individual nodes adopt to access the system.
    \begin{defi}[Access Strategy] An access strategy $S$ consists of an \textit{access rate} $\lambda_S$ and a probability measure $P_S$ on $\mathcal{Q}$, i.e., $\sum_{Q \in \mathcal{Q}} P_S(Q) = 1$. The strategy is \textit{pure} if $P_S(Q) = 1$ for some $Q \in \mathcal{Q}$; otherwise it is \textit{mixed}.
    \label{def:as}
    \end{defi}
    For asymmetric quorum systems, we replace $\mathcal{Q}$ by $\mathcal{Q}^R$ or $\mathcal{Q}^W$, depending on which access operation is under consideration.

    \begin{defi}[Load] The load induced by $S$ on a node $u_i$~is
      \[\ell_S(i) = \sum_{Q \in \mathcal{Q}:u_i \in Q} \lambda_S P_S(Q).\]
    The \textit{system load} induced by $S$ on a quorum system $\mathcal{Q}$ is the maximal load
    induced by $S$ on any node in $\mathcal{U}$, i.e.,
      \[I\!\!L_S(\mathcal{Q}) = \max_{u_i \in \mathcal{U}} \ell_S(i).\]
    \vspace{-2ex}
    \label{def:ld}
    \end{defi}
    Intuitively, this metric measures the evenness of load distribution within the whole system: the lower the system load, the more balanced the load is distributed.

    \subsubsection{Robustness} As another important metric, \textit{robustness} indicates the ability of a quorum system to cope with node failures (viz. its fault tolerance). Many measures have been proposed for this metric, we choose the most straightforward one: the size of the intersection between two quorums.
    \begin{defi}[Robustness] The robustness of a quorum system $\mathcal{Q}$ is the size of the minimum intersection between an arbitrary pair of quorums
      \[I\!\!R(\mathcal{Q}) = \min_{Q_i, Q_j \in \mathcal{Q}} |Q_i \cap Q_j|.\]
    For asymmetric quorum systems, $Q_i, Q_j \in \mathcal{Q}$ is hence replaced by $Q_i \in \mathcal{Q}^W$ and $Q_j \in \mathcal{Q}^R$.
    \label{def:rob}
    \end{defi}
    It is straightforward to see that, if the system robustness is $k$, then any node failures involving less than $k$ nodes will not affect the intersection property of the system.

  \subsection{Related Work on Conventional Quorum Systems}
\begin{figure}[htb]
        \begin{center}
        \includegraphics[width=.8\columnwidth]{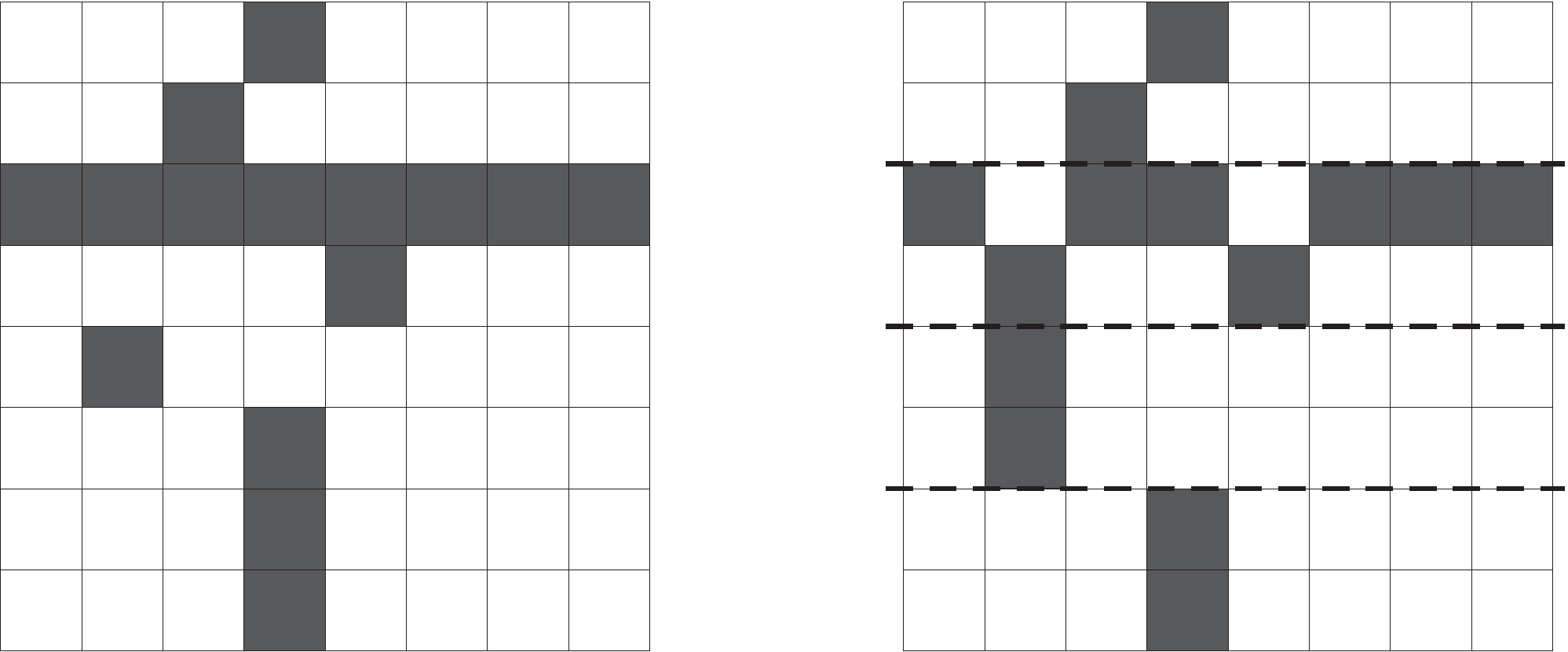}
        \parbox{\columnwidth}{\center\parbox{.42\columnwidth}{\center\scriptsize(a) Grid quorum}
                             \parbox{.48\columnwidth}{\center\scriptsize(b) B-Grid}}
        \caption{Quorums system based on 2D linear curves.}
        \label{fig:grid}
        \end{center}
        \vspace{-2ex}
    \end{figure}

    Traditional quorum systems are confined in 2D space, and hence only allow for limited designs, such as the grid shown in Figure~\ref{fig:grid}(a),
    or the B-Grid \cite{NaorW-SIAMJC98}, shown in Figure~\ref{fig:grid}(b), for improving the robustness. Similar ideas were re-introduced into \textit{mobile ad hoc networks} (MANETs) and WSNs \cite{HaasL-ToN99,YeLCLZ-MobiCom02,LiuHZ-Sensys04}, though sometimes under different names. These designs are often so rigid that they allow very little tunability that adapts a system to various application requirements.

    To improve the flexibility of the quorum systems, \textit{probabilistic quorum systems} \cite{MalkhiRW-IC01} were introduced to relax the intersection rule (making it a random variable) and to leave more freedom in trading load for robustness; they were later applied to MANETs to cope with node mobility \cite{HaasL-ICC99}. Interested readers are referred to \cite{LuoHE-MobiHoc03,LuoEH-TMC04,FriedmanKA-DSN08} for more recent developments in probabilistic quorum systems. In general, probabilistic quorum systems are designed to cope with system dynamics (e.g., node mobility), hence they are trading system efficiency for higher robustness. As we explained in Sec.~\ref{sec:intro}, energy efficiency is a crucial issue in WSNs, whereas nodes in WSNs are often static. Consequently, we advocate a deterministic design for quorum systems, while relying on other techniques (rather than pure randomization) to improve its flexibility.

  \subsection{Quorum Systems in A Projective Space} \label{sec:qsps}
    Recently, a new design methodology for (deterministic) quorum systems was proposed in~\cite{SarkarZG-MobiCom06}. This method suggests using projective map to first ``lift'' the 2D network area onto a 3D surface, a sphere, then design quorum systems on the 3D surface, and finally project the designed system back to the 2D area. As the system design is done in the 3D space, it allows much more diversity in ``shaping'' the quorums, and hence has a potential to deliver more flexible system designs. The practicality of this design approach is backed by the \textit{trajectory based forwarding}~\cite{NiculescuN-MobiCom03}, which allows a routing path to be defined by a continuous curve.

\begin{figure}[htb]
      \begin{center}
      \includegraphics[width=.8\columnwidth]{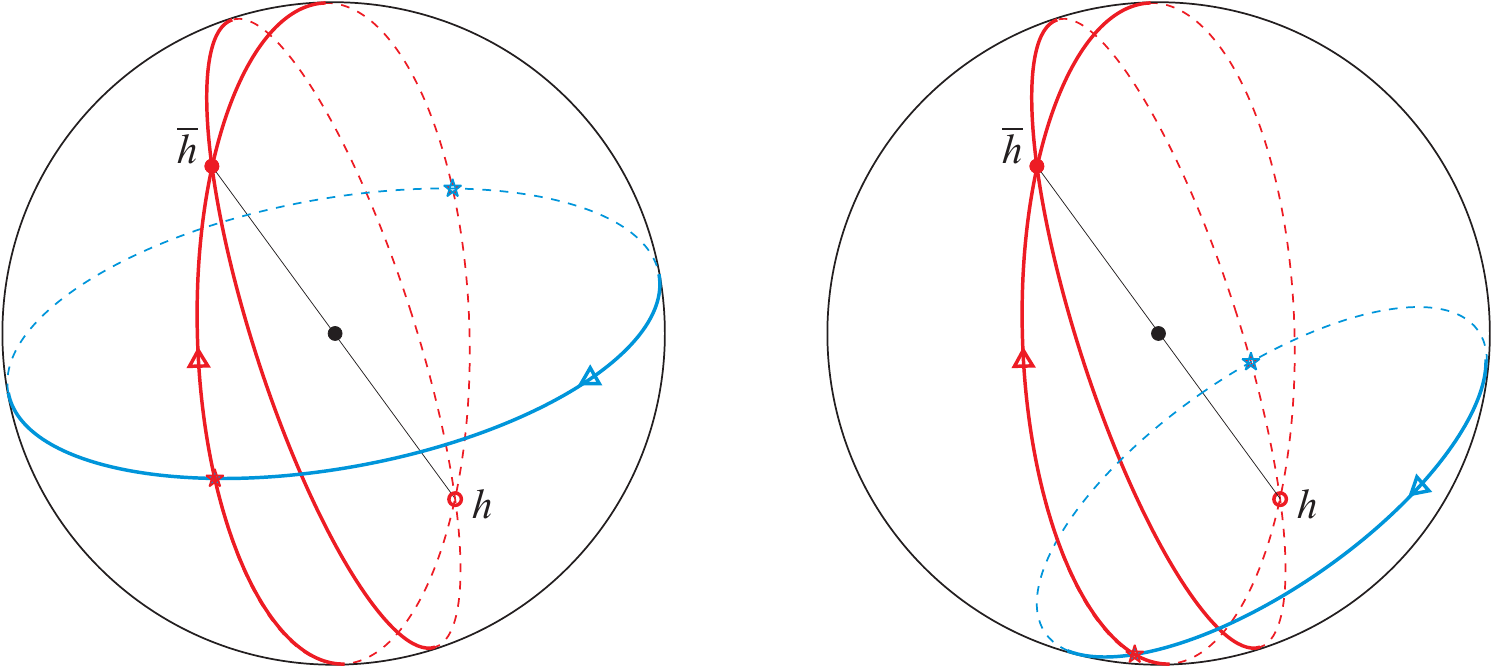}
      \parbox{\columnwidth}{\parbox{.49\columnwidth}{\center\scriptsize(a) Symmetric quorum system $\mathcal{Q}_G$}
                             \parbox{.49\columnwidth}{\center\scriptsize(b) Asymmetric quorum system $\mathcal{Q}_L$}}
      \caption{Quorum systems designed in~\cite{SarkarZG-MobiCom06}. We use red (resp. blue) color to indicate quorums accessed by a write (resp. read) access. For quorums in red, the corresponding geographical hash location $h$ and its antipodal point $\bar{h}$ are shown. We also use pentagrams to represent the intersection between quorums, and triangles to represent the nodes that access a quorum.}
      \label{fig:drulings}
      \end{center}
      \vspace{-3ex}
    \end{figure}

    Given a certain data type, two designs are proposed in~\cite{SarkarZG-MobiCom06}.\footnote{In the original paper, a quorum system design is termed a \textit{double ruling scheme}. The two designs we discuss here are named \textit{double rulings retrieval} and \textit{distance-sensitive retrieval}.}
    \begin{itemize}
      \item [$\mathcal{Q}_G$] Symmetric quorum systems with each quorum represented by a great circle. The access strategy for a write access is pure as the corresponding great circle is fixed by two points: the node accesses a quorum and the geographical hash $h$ of the data type.
      \item [$\mathcal{Q}_L$] Asymmetric quorum systems with write quorums represented by great circles and read quorums by latitude circles. While the access strategy for a write access is the same as the first design, that of a read access also becomes pure, as the circle of each read quorum is also defined by the node access the quorum and the geographical hash $h$ of the data type.
    \end{itemize}
    We illustrate the two designs in Fig.~\ref{fig:drulings}.

    The projective map used in~\cite{SarkarZG-MobiCom06} is stereographic projection, which is easy to use due to its closed-form expression. However, using stereographic projection, the north pole is mapped to infinity on the plane. Consequently, as the sphere cannot be fully covered by the (mapped) network area, conclusions drawn through geometric analysis on the sphere surface may not be valid for the original 2D network area. For example, two curves intersecting on the sphere may not retain the property within the network area, as the intersection may be out of the boundary: only one of the two intersections between two great circles is guaranteed in \cite{SarkarZG-MobiCom06}.

    In terms of quorum system design, apart from presenting heuristics, no rigorous definitions and metrics are provided for the quorum systems, thus no formal analysis is given in \cite{SarkarZG-MobiCom06} to evaluation the performance of the designed system. Also, only planar curves are used to represent quorums on the 3D surface, which significantly confines the design flexibility. In addition, as we will show later, any two planar curves can intersect at up to two points on a sphere, the system robustness is hence fixed and cannot be tuned.

\section{Geometric Quorum Systems for Data Access} \label{sec:gqs}
  In this section, we first introduce the geometry background and define our network model, along with the properties and metrics of geometric quorum systems (GQS). Then we analyze the performance of existing designs, based on the defined metrics. Finally, we present our asymmetric quorum systems, GeoQuorum, that makes use of spatial curves to substantially improve the flexibility in fine-tuning system performance.

  \subsection{Background on Computational Conformal Geometry} \label{sec:bkgccg}
    \textit{Computational conformal geometry} (CCG) is an emerging research field spanning computer science and pure mathematics. It focuses on developing the computational methodologies on discrete surfaces to discover conformal geometric invariants. Intuitively speaking, a conformal map is a function that preserves the angles. Due to its shape preserving properties, computational conformal geometry has broad applications in both pure theoretic research, such as mathematics, theoretical physics, and engineering applications, such as computer graphics, computer-aided design, computer vision, etc. See \cite{Gu09} for a recent survey of CCG.

\begin{figure*}[htb]
      \begin{center}
      \parbox{\textwidth}{\parbox{.195\textwidth}{\center\includegraphics[width=.195\textwidth]{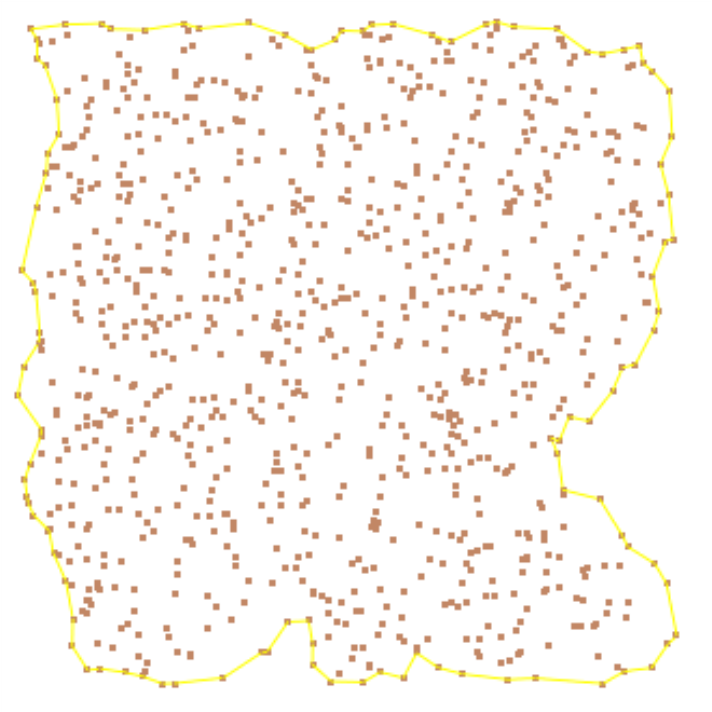}}
                          \parbox{.195\textwidth}{\center\includegraphics[width=.195\textwidth]{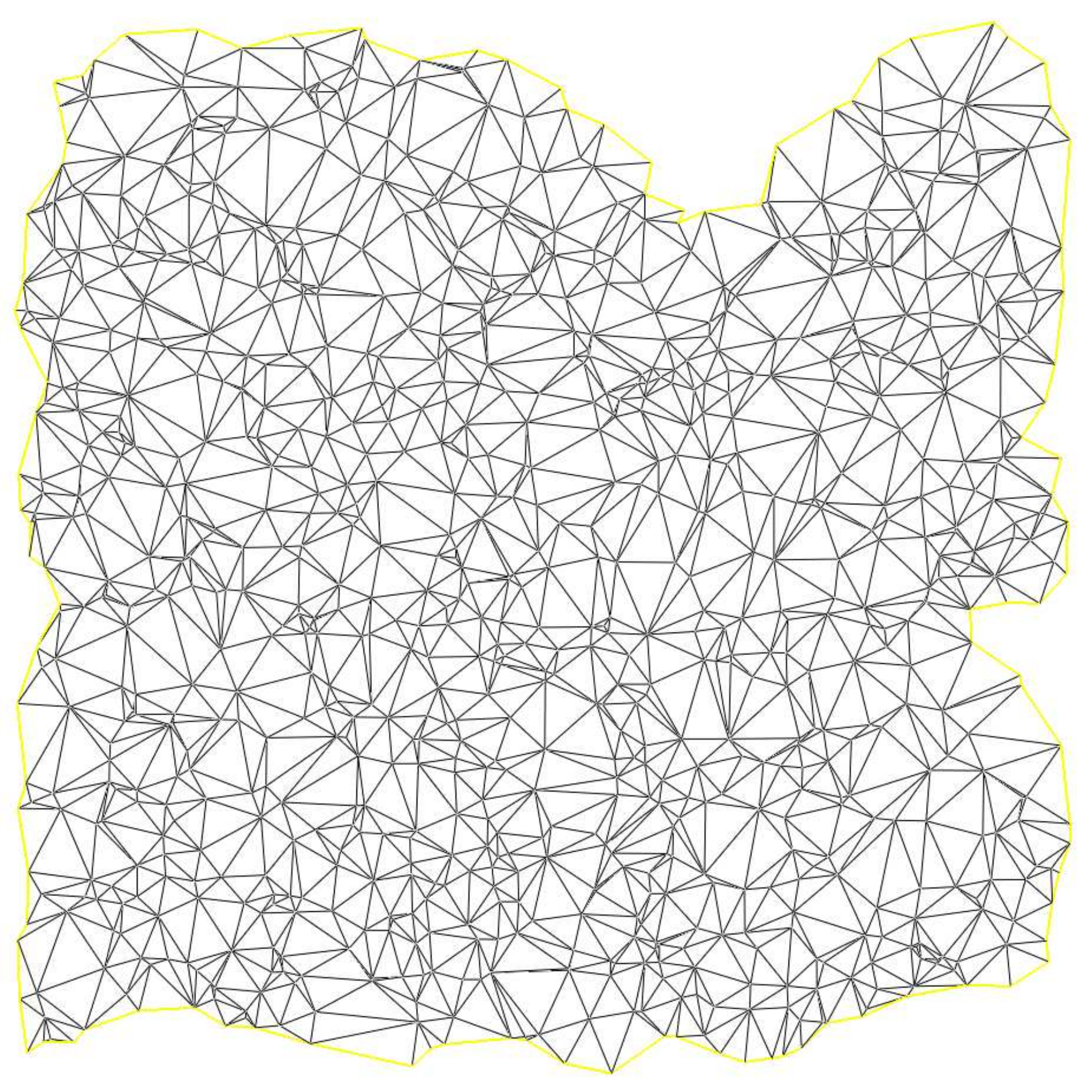}}
                          \parbox{.195\textwidth}{\center\includegraphics[width=.195\textwidth]{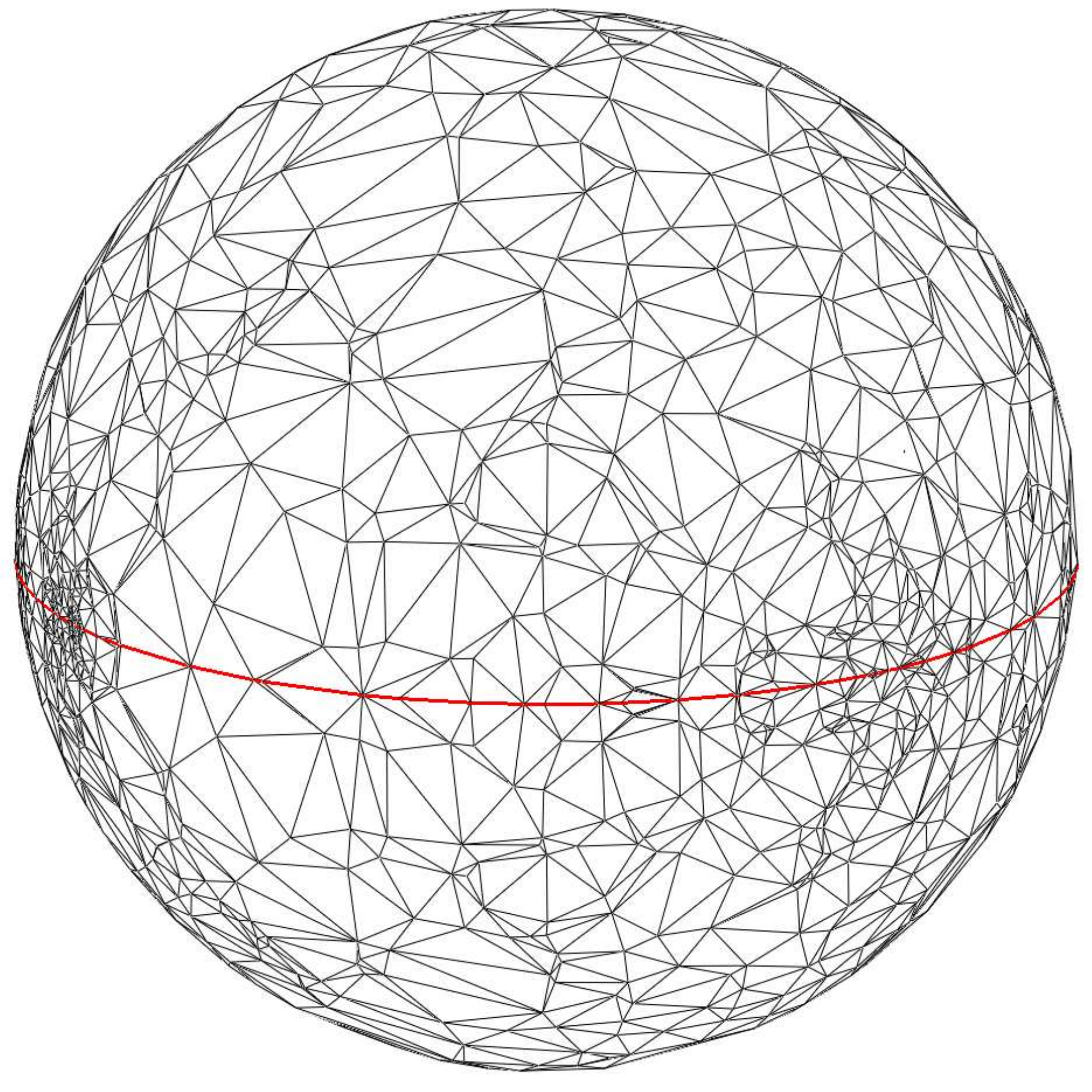}}
                          \parbox{.195\textwidth}{\center\includegraphics[width=.195\textwidth]{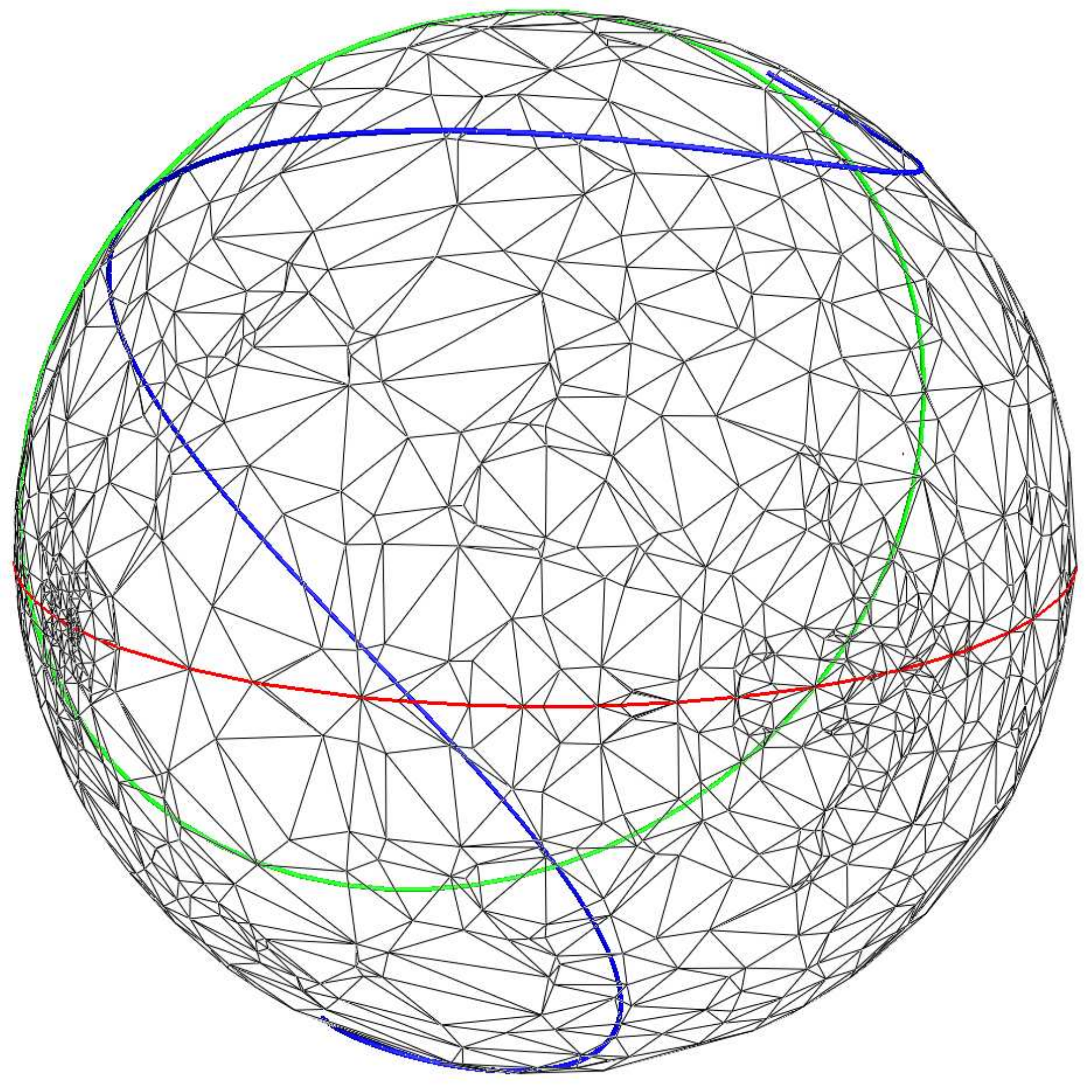}}
                          \parbox{.195\textwidth}{\center\includegraphics[width=.195\textwidth]{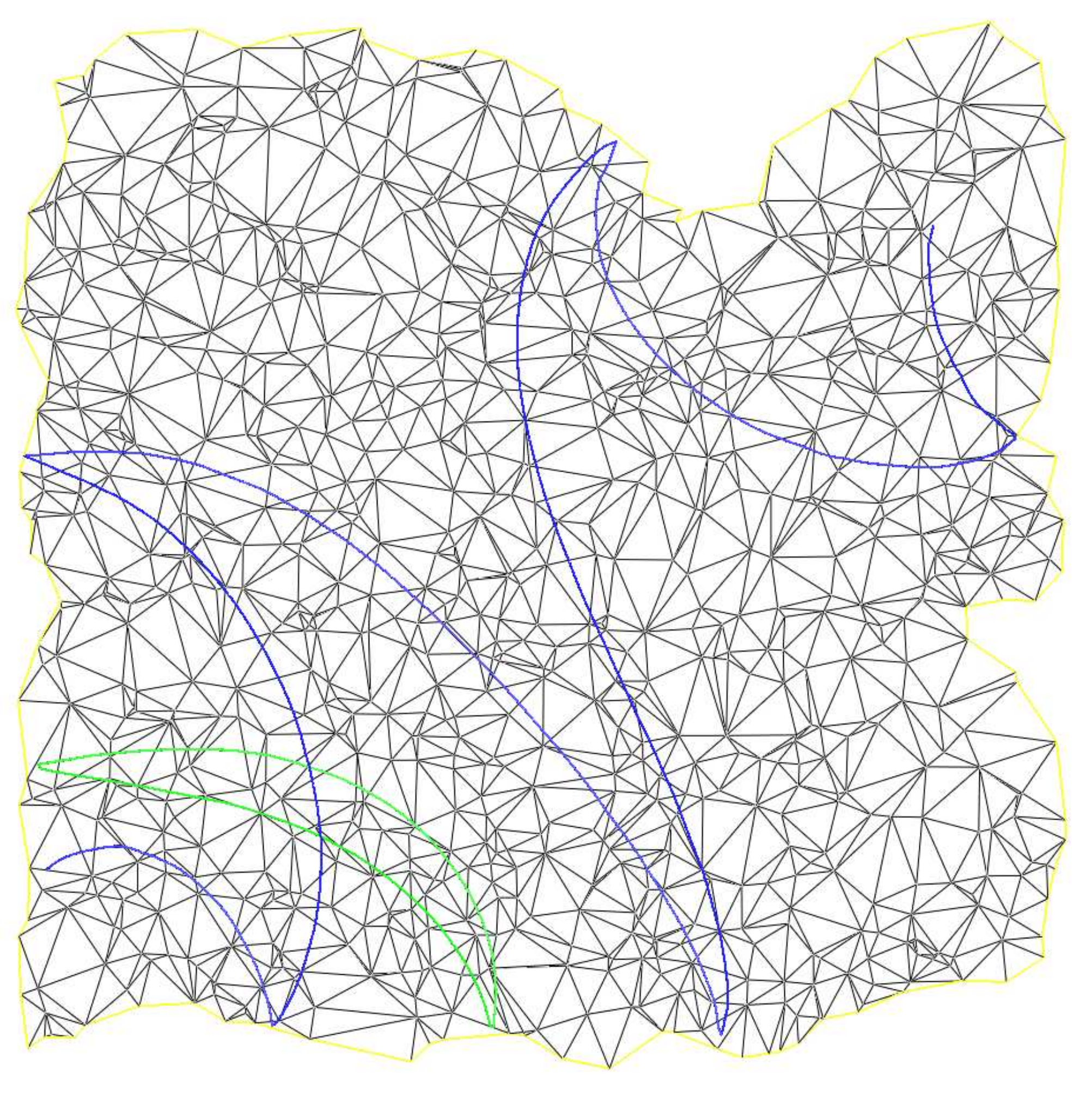}}}
      \parbox{\textwidth}{\parbox{.195\textwidth}{\center\scriptsize(a)}
                          \parbox{.195\textwidth}{\center\scriptsize(b)}
                          \parbox{.195\textwidth}{\center\scriptsize(c)}
                          \parbox{.195\textwidth}{\center\scriptsize(d)}
                          \parbox{.195\textwidth}{\center\scriptsize(e)}}
      \caption{Taking an arbitrary simply connected region $D$ and a node set inside it as input (a), our design starts with a Delaunay triangulation and a construction of a closed genus-0 surface $\bar D$ by double covering (b). Then $\bar D$ is conformally mapped to the unit sphere, such that the original boundary $\partial D$ is the equator, the red circle in (c). Note that the input region $D$ and its copy $D'$ are mirror reflected with respect to the equator. This design allows us to use various spherical curves to form quorums, such as circles and spirals shown in (d) and their inverse-map in $D$ (e).}
      \label{fig:mapping}
      \end{center}
      \vspace{-2ex}
    \end{figure*}

    Here we briefly discuss the CCG tools that we will use in this paper, which is also illustrated in Fig.~\ref{fig:mapping}.
    Given a simply connected shape $D\in\mathbb{R}^2$ with boundary $\partial D$ and a node set inside it, we first construct a Delaunay triangulation using~\cite{Shewchuk96}, then we construct a closed surface by double covering: make a copy of $D$, denoted by $D'$, reverse its orientation and glue $D$ and $D'$ along the common boundary. The resulted surface $\bar D$ is a closed surface of genus 0, i.e., topologically equivalent to a sphere. We then compute a harmonic function $\phi$ mapping $\bar D$ to the unit sphere, i.e., $\phi:\bar{D}\rightarrow\mathbb{S}^2$ such that $\bigtriangleup\phi=0$ where $\bigtriangleup$ is the Laplace-Beltrami operator. This map has the following promising properties:
    \begin{itemize}
      \item $\phi$ is conformal, thus, there is no angle distortion~\cite{DBLP:journals/tmi/GuWCTY04};
      \item Due to symmetry, $D$ and $D'$ are mapped to the north and south hemispheres respectively, i.e., $\phi(D)$ and $\phi(D')$ are mirror reflected with respect to the equator.
      \item  The boundary $\partial D$ is mapped to the equator. As $\partial D$ could be arbitrary 2D simple curve, this algorithm applies for arbitrary 2D simply connected region $D$.
    \end{itemize}

    Here we should emphasize that there are other ways to compute the conformal mappings on the sphere, such as the stereographic projection used in~\cite{SarkarZG-MobiCom06}. However, the harmonic map based method as mentioned above allows us to map arbitrary simply connected region to cover the whole sphere, thus avoiding various issues involved in stereographic projection.

  \subsection{Network Model and Geometric Design Basics} \label{sec:gqsmd}
    We represent a WSN by $\mathcal{U}$, with $u_i \in \mathcal{U}$ being a sensor node. $\mathcal{U}$ also serves as the universe upon which a quorum system can be defined.

    \subsubsection{Geometric Model of WSNs} We apply the tool discussed in Sec.~\ref{sec:bkgccg} to map the network area to a sphere of \textbf{unit radius}. For the reverse projection, any curve that passes across the equator has its upper and lower sections projected separately to the two network areas. Then two (projected) sections are combined to get the projection on the original network area, as shown in Fig.~\ref{fig:mapping}(e).
    %
    %
    This improved map allows us to perform geometric analysis on the whole sphere surface.

    In order to facilitate the analysis of geometric quorum systems in Sec.~\ref{sec:gqsals}, we assume that the tessellation on the sphere surface is regular and use the vertices to represent sensor nodes.
    %
    %
    Such a model makes sense in dense WSNs where nodes are uniformly distributed; the tessellation vertices can be considered as representatives of the nodes in a close-by region. We use this model only to simplify analysis, our numerical simulations still take arbitrarily deployed WSNs as input, and the tessellation on a sphere simply results from the triangulation of the network nodes, as shown in Fig.~\ref{fig:mapping}(c).

    \subsubsection{Geometric Quorum Systems} We extend the conventional definitions for quorum systems (presented in Sec.~\ref{sec:qsbs}) to geometric quorums system (GQS).
    \begin{defi}[Geometric Quorum System] ~~A GQS $\mathcal{Q}$ is a set of curves in space $\mathcal{A}$ ($\mathcal{U} \subset \mathcal{A}$), such that every two curves intersect. Each curve in $\mathcal{Q}$ defines a quorum.
    \end{defi}
    The definition for asymmetric quorum systems is omitted; one simply splits $\mathcal{Q}$ into $\mathcal{Q}^W$ and $\mathcal{Q}^R$, and intersection is only required between the two sets. We keep using the same definition for access strategy (\textit{Definition~\ref{def:as}}), load (\textit{Definition~\ref{def:ld}}), and robustness (\textit{Definition~\ref{def:rob}}), but the interpretations are slightly different. In particular, a sensor node $u_i \in \mathcal{U}$ is also a tessellation vertex of $\mathcal{A}$, and $u_i \in Q$ means $u_i$ is a vertex of a triangle passed by the curve $Q$. The system load defined for a WSN is the maximum energy consumption for a certain tessellation vertex to \textbf{transmit} the data (for write) or queries (for read) to the quorum at which they aim, representing the load balancing effect of the quorum system.

    Unlike traditional distributed systems, the energy efficiency (or total energy consumption of the whole WSN) is also a major concern of WSNs. Let $M(Q)=|\{u \in \mathcal{A}|u \in Q\}|$ be a measure of the total energy consumption of a quorum $Q$, we further define a metric to measure this performance aspect.
    \begin{defi}[Total Load] The total load induced by $S$ on a certain quorum $\mathcal{Q}$ is
      \[I\!\!L_T(\mathcal{Q}) = \sum_{Q \in \mathcal{Q}} \lambda_S P_S(Q) M(Q).\]
    \label{def:tld}
    \end{defi}
    In general, each node may take a different access strategy. To simplify the analysis, we only distinguish between two types of strategies, namely $S_R$ and $S_W$ for read and write respectively.

  \subsection{Existing Quorum System Designs} \label{sec:gqsals}
    In this section, we analyze the performance of the two quorum systems designed in \cite{SarkarZG-MobiCom06}, namely $\mathcal{Q}_G$ and $\mathcal{Q}_L$ described in Sec.~\ref{sec:qsps}. We will point out that, if we design quorum systems based only on planar curves, the system robustness is very limited and the system load can be very high. More importantly, such a design may be lack of flexibility to cope with a high asymmetry between $S_R$ and $S_W$, which can lead to an unnecessarily high total load. Based on our analysis, we also propose certain remedies to improve these two designs.

    \subsubsection{Symmetric Quorum Based on Great Circles ($\mathcal{Q}_G$)} We first show that \textbf{the system robustness is limited}. Instead of considering only great circles, the following result encompasses all possible planar curves on a sphere.
    \begin{prop} Any two distinct planar curves intersect at most at two points on a sphere.
    \end{prop}
    \begin{IEEEproof} Any planar curve on the sphere surface is the intersection between the sphere surface and a certain plane (\textit{cutting plane} hereafter), and two planar curves intersect each other iff their corresponding cutting planes intersect each other. In addition, the intersection of two planar curves is a subset of that of the two cutting planes. Now, as the intersection of two cutting planes (of two distinct planar curves) is bounded to be a line, and as a line may intersect the sphere surface at (at most) two points, the proposition follows.
    \end{IEEEproof}
    The following corollary is immediate.
    \begin{coro} The robustness of $\mathcal{Q}_G$ is no more than 2.
    \end{coro}
    The implication of this result is very clear: if two nodes of a WSN fail, the intersection property of the quorum system may be violated. If we consider a specific pair of quorums, the probability of the two intersecting nodes fail simultaneously is not negligible. Moreover, the robustness is fixed no matter what planar curves are used to define a pair of quorums.

    Secondly, we demonstrate that the access strategy defined by geographical hash leads to \textbf{very unbalanced load distribution}. It is straightforward to see that the write strategy is pure. In particular, the quorum that can be chosen by a writing node is fixed by that node and the hash location $h$.
    \begin{prop} \label{prp:load} Let $\mathcal{D} = \{d_1, d_2, \cdots, \}$ be a set of data types in a WSN, and let $N(d)$ be the number of nodes contributing to a data type $d \in \mathcal{D}$. We have
      \[I\!\!L_S(\mathcal{Q}_G) = \lambda_{S_W} \max_{d \in \mathcal{D}} N(d) + \lambda_{S_R} \mathcal{O}(1)\]
    \end{prop}
    \begin{IEEEproof} The proof is based on \textit{Definition~\ref{def:ld}}. The first term is incurred by the write access. As the write strategy is pure, we have $\ell_{S_W}(i) = \sum_{Q \in \mathcal{Q}:u_i \in Q} \lambda_{S_W} \mathbf{1}_Q$. Therefore, the load is computed by counting how many quorums include a node $u_i$. Because every write quorum for a given data type includes the corresponding hash location, the maximum loaded node is at the hash location whose corresponding data type has the largest number of contributors. The second term is introduced by the read access. As the read strategy can be mixed, the optimal strategy is apparently a uniform distribution over all possible quorums, given the assumption of a regular tessellation on the sphere surface. Consequently, the load introduced by a read access is identical for all nodes and can be bounded by a constant (which depends on the density of the tessellation vertices).
    \end{IEEEproof}
    Note that the first term is fully determined by the nature of the sensory data produced by a WSN. In the case that many nodes are contributing to the same data type, the system load can be extremely high.

\begin{figure}[htb]
    \begin{center}
    \includegraphics[width=\columnwidth]{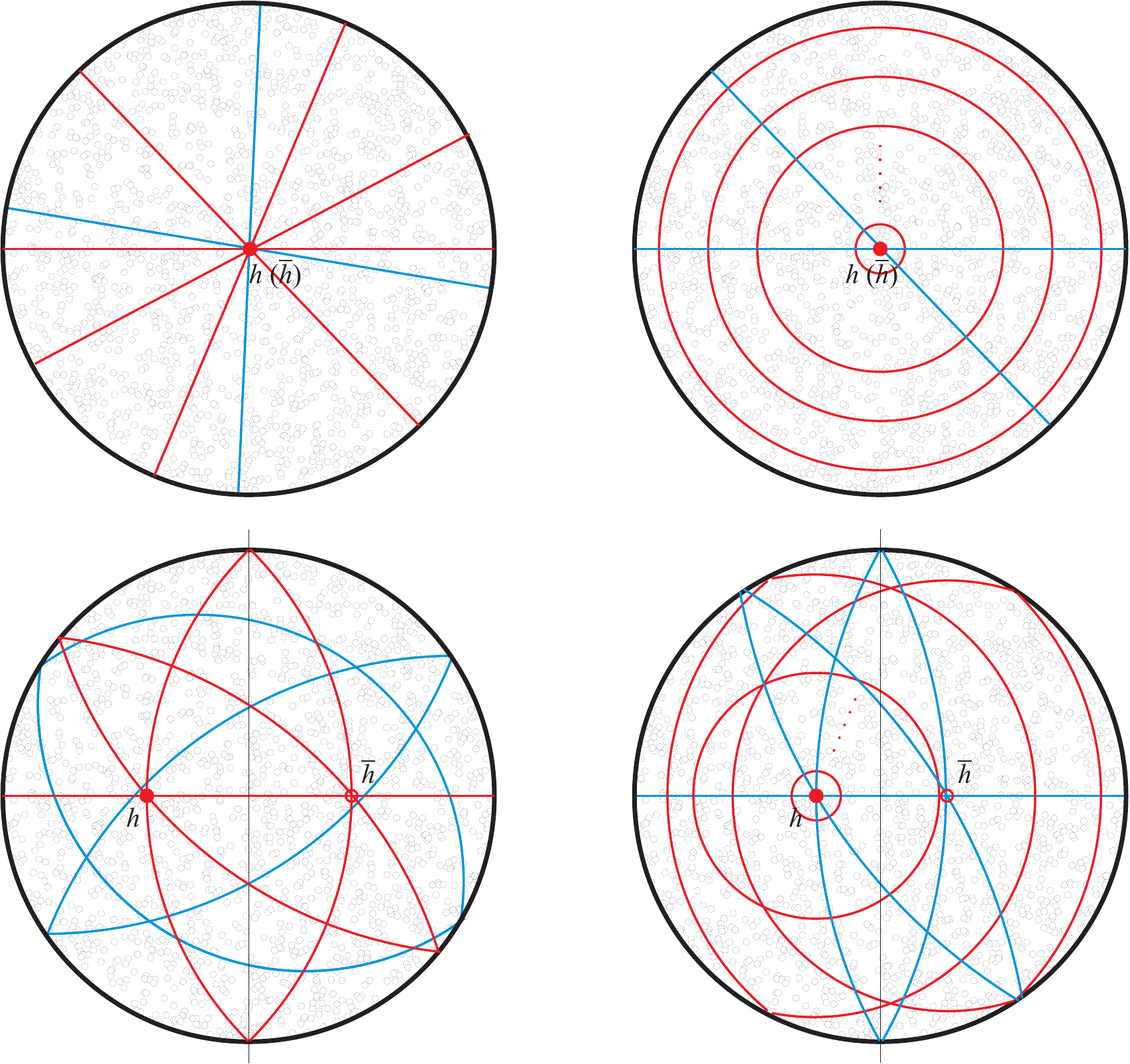}
    \parbox{\columnwidth}{\parbox{.49\columnwidth}{\scriptsize(a) Symmetric quorum systems $\mathcal{Q}_G$}
                             \parbox{.49\columnwidth}{\hfill\scriptsize(b) Asymmetric quorum systems $\mathcal{Q}_L^d$}}
    \caption{Geometric quorum systems designed using planar curves in 3D. We mark write (reps. read) quorums by red (resp. blue) color, and we also show the corresponding geographical hash location $h$ and its antipodal point $\bar{h}$.}
    \label{fig:existing}
    \end{center}
    \vspace{-1ex}
    \end{figure}

    In practice, this result suggests that, for every node contributing to a certain data type, a certain amount of communication load is imposed upon the nodes around the hash location at a rate $\lambda_{S_W}$. Therefore, the more nodes contributing to this data type, the sooner the nodes around the hash location will run out of their energy storage. We show the two sets of quorums in Fig.~\ref{fig:existing}(a); we plot the 2D project in the network area to avoid confusion in 3D representations. One extreme case is shown in the top figure, where the hash location coincides the network center,
    imposing very high load there.\footnote{This actually brings us back to the distributed hash tables (GHTs) based scheme \cite{RatnasamyKYYEGS-WSNA02}, totally annihilating the benefit of quorum systems.}

    One possible \textbf{remedy} $\mathcal{Q}_G^m$ is to adopt a mixed strategy for write access. For example, it is straightforward to see that, if the nodes contributing to a certain data type are uniformly distributed, a mixed strategy with uniform distribution among all possible quorums (great circles) may balance the load. However, as great circle is the longest planar curve on a sphere, always using it as quorum can lead to unnecessarily high total load, in particular if, for example, the read and write strategies have very different access rates.\footnote{In reality, the popularity of different data types may vary a lot, this may further differentiate the access rates. Therefore, in order to cope with the variety of the access rates, we need a better quorum system design to fully utilize the freedom in the 3D space.}

    \subsubsection{Asymmetric Quorum Based on Great Circles and Latitude Curves ($\mathcal{Q}_L$)}
    According to Sec.~\ref{sec:qsps}, $\mathcal{Q}_L$ shares the same write quorums with $\mathcal{Q}_G$. Therefore, the same drawbacks, namely low robustness and high system load, persist. Actually, what $\mathcal{Q}_L$ improves (comparing with $\mathcal{Q}_G$) is total load, as shown by the following proposition.
    \begin{prop} For a given data type, if the nodes that access (by read or write) a quorum system are uniformly distributed on the sphere, the part of the total load contributed by the read access of $\mathcal{Q}_L$, $I\!\!L_T^R(\mathcal{Q}_L)$, is $\frac{\pi}{4}$ of that of $\mathcal{Q}_G$.
    \end{prop}
    \begin{IEEEproof} Let $R_G$ and $R_L$ be the radius of a great circle and a latitude circle, respectively. Based on the assumption of a regular tessellation on the sphere surface, we use the length of a quorum $Q$ (a curve) to represent $M(Q)$, the measure of its total energy consumption. As shown in Fig.~\ref{fig:latitude},
    \begin{figure}[htb]
    \begin{center}
    \includegraphics[width=.4\columnwidth]{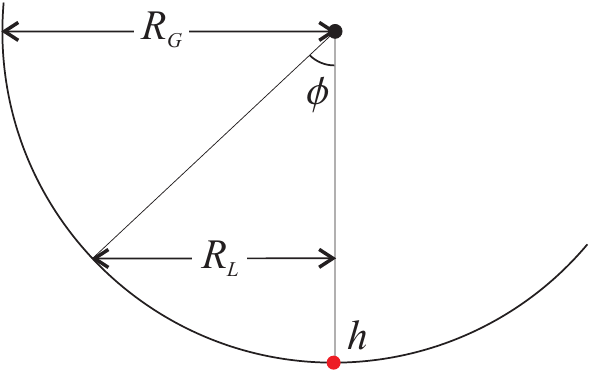}
    \caption{The relation between $R_G$ and $R_L$.}
    \label{fig:latitude}
    \end{center}
    \end{figure}
    $R_L$ is determined by $R_G$ and the angle $\phi$. Hence we have
        \begin{eqnarray}
        \mathbf{E}_\phi\left(M(Q_L)\right) &=& 2 \pi \int_{0}^{\frac{\pi}{2}} \mathrm{Pr}\{R_L = R_G\sin(\phi)\} R_L \nonumber \\
        &=& 2 \pi R_G \int_{0}^{\frac{\pi}{2}} \sin^2(\phi) d\phi ~~=~~ \frac{\pi^2}{2} R_G \nonumber
        \end{eqnarray}
    Since we have the total load contributed by a write quorum of $\mathcal{Q}_G$ as $2 \pi R_G$, the ratio $\frac{\pi}{4}$ follows.
    \end{IEEEproof}
    As the part of the total load contributed by the write access is the same for $\mathcal{Q}_G$ and $\mathcal{Q}_L$, the above result shows that load reduction brought by $\mathcal{Q}_L$ may become marginal if a lot of nodes are contributing to a certain data type.

    One immediate \textbf{remedy} to cut down both the system load and total load is $\mathcal{Q}_L^d$, a dual design of $\mathcal{Q}_L$. In other words, we swap the quorums for read and write, with read quorums given by great circles and write quorums given by latitude circles, as illustrated in Fig.~\ref{fig:existing}(b). This design can reduce the system and total load if only a few write quorums are contributing to a data type, because, though all the read quorums for a certain data type pass through the hash location and its antipodal point, the read access may terminate (i.e., reaches an intersection with some write quorum) before reaching those locations. Unfortunately, the load reduction effect diminishes with an increasing number of write quorums.

  \subsection{GeoQuorum: System with Spatial Quorums} \label{sec:gqsgq}
    Besides the various drawbacks we have pointed out, a main disadvantage of $\mathcal{Q}_G$ and $\mathcal{Q}_L$ is the lack of flexibility to be fine-tuned, hence they cannot adapt to different access rates. In this section, we present GeoQuorum as a new design. GeoQuorum makes use of spatial curves to form quorums, hence allows a great deal of freedom in fine-tuning the system performance.

    GeoQuorum is an asymmetric quorum system, with write and read quorums formed by different type of curves. Specifically, we have
    \begin{itemize}
    \item \textbf{write quorums} are formed by circles with adjustable radius $R_W$, where $R_W$ can be tuned according to the access rate.
    \item \textbf{read quorums} are formed by a special spherical spiral, defined as follows:
        \begin{eqnarray}
        x &=& \cos \left(\theta + \theta_0 \right) \cos \phi \\
        y &=& \sin \left(\theta + \theta_0 \right) \cos \phi \\
        z &=& \sin \phi
        \end{eqnarray}
        where $\phi = a\theta$ and $\phi \in \left[-\frac{\pi}{2},\frac{\pi}{2}\right]$. Let $\alpha$ be the angle (with respect to the sphere center) between two consecutive loops ($\Delta \theta = 2\pi$), we have $\alpha = 2a\pi$. The parameter $a$ is determined by $R_W$ and the required robustness.
      \item \textbf{access strategy} is mixed:
      \begin{itemize}
        \item a \textbf{write quorum} is randomly chosen among all circles passing through the node that executes a write access, and
        \item a \textbf{read quorum} starts from the node that executes a read access and ends at its antipodal point, with a randomly chosen $\theta_0$.
      \end{itemize}
    \end{itemize}

\begin{figure}[htb]
    \begin{center}
        \includegraphics[width=.9\columnwidth]{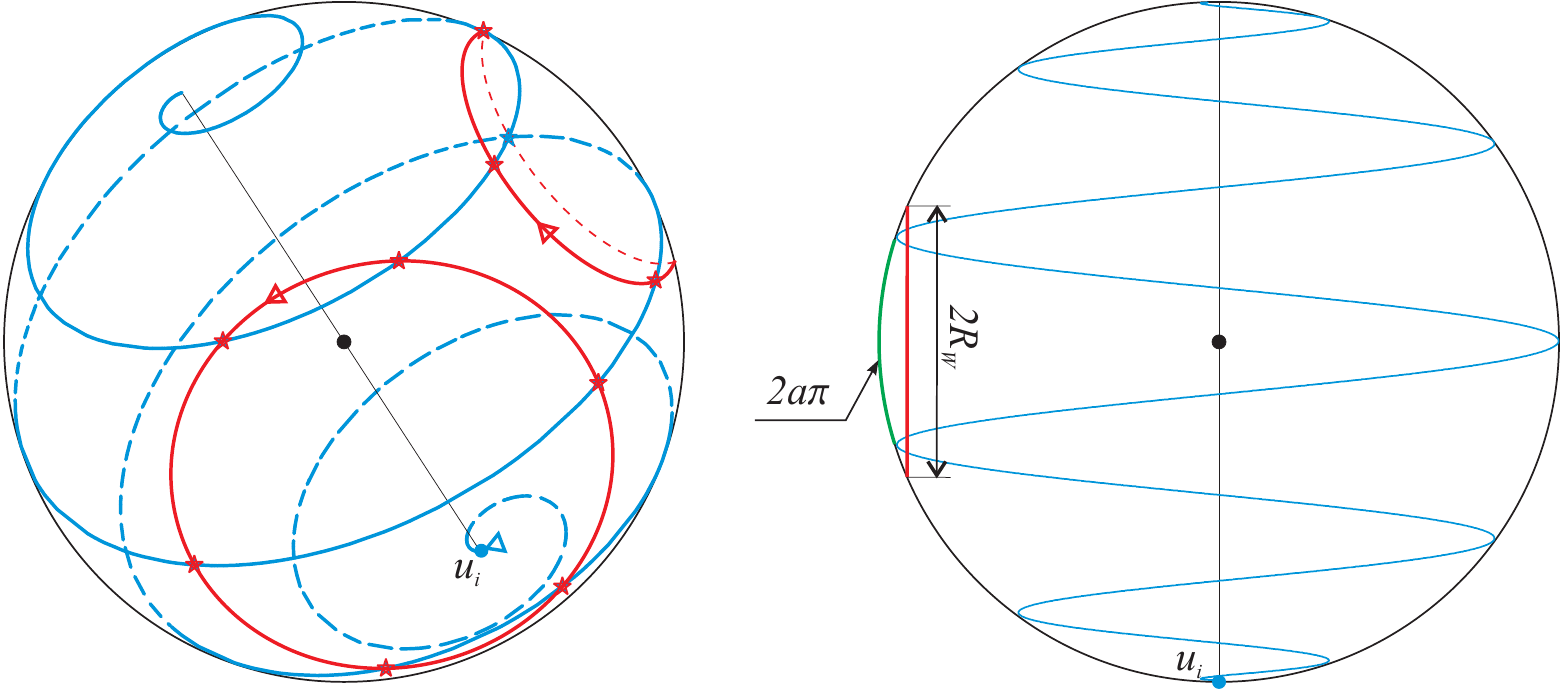}
        \parbox{\columnwidth}{\parbox{.49\columnwidth}{\center\scriptsize(a) GeoQuorum}
                             \parbox{.49\columnwidth}{\center\scriptsize(b) Proof of \textit{Proposition~\ref{prp:rbst}}}}
    \caption{Geometric quorum system designed using spatial curves in 3D.}
    \label{fig:spiral}
    \end{center}
    \end{figure}

    We illustrate such a quorum system in Fig.~\ref{fig:spiral}(a).
    Note that the current design is based on the assumption that $\lambda_W > \lambda_R$; otherwise we adopt a dual design where we swap the write and read quorums. We first show the relation between $R_W$ and $a$ by the following proposition.
    \begin{prop} If $R_W \ge ka\pi, a \in (0,0.5)$, then the robustness of GeoQuorum is at least $2k$. \label{prp:rbst}
    \end{prop}
    \begin{IEEEproof} As shown in Fig.~\ref{fig:spiral}(b), if $2R_W \ge 2a\pi$ (i.e., $k=1$, the arc length delimited by the write quorum (a circle) is bounded to be larger than $2a\pi$, which is indeed the distance between two consecutive loops of the spherical spiral on the sphere surface. Consequently, the read quorum formed by the spherical curve has a least one loop passing through the write quorum, resulting in 2 intersections. Every time $k$ is increased by one (i.e., $2R_W \ge 2ka\pi$), an extra loop of the spiral passes the circle, given 2 more intersections, hence $2k$ in total.
    \end{IEEEproof}
    Therefore, given a certain robustness requirement, we have a one-to-one correspondence between $R_W$ and $a$: $R_W = ka\pi$, as choosing the smallest circle minimizes the incurred system and total load. Under the assumption that $\lambda_W > \lambda_R$, we may choose to tune $R_W$ according to $\lambda_W$ (the higher the rate the smaller $R_W$ is), then we match $a$ to $R_W$ based on the required robustness. Due to the use of mixed access strategy and the parameterized design, GeoQuorum can be tailored to meet the application requirements, such that both system load and total load can be reduced; which we will show in Sec.~\ref{sec:comp}.

    Interestingly, our design includes $\mathcal{Q}_G$ and $\mathcal{Q}_L$ as special cases. The following proposition shows that, by making $a$ sufficiently large, the spherical spiral becomes a great circle.
    \begin{prop} If $a \rightarrow \infty$, the spiral curves becomes part of a great circle, whose orientation is determined by $\theta_0$.
    \end{prop}
    \begin{proof} Let $\mathbf{C}(\theta)=(x(\theta,\phi), y(\theta,\phi), z(\theta,\phi))^{T}$ denote the spherical spiral curve, where $\phi=a\theta$. A simple computation shows that $\mathbf{C}''=-a^2(\mathbf{C}+\mathbf{D})$, where $\mathbf{D}=(d_x, d_y, 0)^T$ and $d_x=\frac{-1}{a^2}(\cos(\theta+\theta_0)\cos(a\theta)-2a\sin(\theta+\theta_0)\sin(a\theta))$, $d_y=\frac{-1}{a^2}(\sin(\theta+\theta_0)\cos(a\theta)+2a\cos(\theta+\theta_0)\sin(a\theta))$.
    As $\lim_{a\rightarrow\infty}\mathbf{D}=\mathbf{0}$, the curvature vector of $C$ is parallel to the normal vector of sphere when $a\rightarrow\infty$, i.e., the geodesic curvature of $C$ is zero. Observe that the geodesics on the sphere are great circles, so $\bf C$ is part of a great circle. Furthermore, the orientation of the great circle is determined by its binormal vector $\mathbf{C}'\times\mathbf{C}''$, which is parallel to $\mathbf{C}'\times\mathbf{C}$ when $a\rightarrow\infty$, but we also have
    $\lim_{a\rightarrow\infty}\frac{\mathbf{C}'\times\mathbf{C}}{\|\mathbf{C}'\times\mathbf{C}\|}=(\sin\theta_{0},\cos\theta_{0},0)^T$,
    hence the proposition follows.
    \end{proof}
    For $a \ge 0.5$, $R_W$ cannot be futher increased. Consequently, the range of $\phi$ has to be extended to $(-\frac{\pi}{2},\frac{3\pi}{2})$ to maintain the robustness. Therefore, if $a \rightarrow \infty$, GeoQuorum shares the same write quorum with $\mathcal{Q}_G$ and $\mathcal{Q}_L$: the great circles.

\section{Simulations} \label{sec:sim}
  We hereby use simulation results to confirm the advantages of GeoQuorum over the existing designs, and also to demonstrate GeoQuroum's ability of fine-tuning load and robustness.

  \subsection{Simulation Settings}
    We randomly put nodes in an area (to be specified for each set of simulations). Then we use Delaunay triangulation to generate the connectivity graph, assuming that the sensor nodes are power controlled such that two nodes are connected by a wireless link iff there exists an edge (of the Delaunay triangulation) between them. We assume that the trajectory based forwarding~\cite{NiculescuN-MobiCom03} is used to guide both write and read accesses, based on the corresponding curves that form the write and read quorums. As far as a curve passes through a triangle, all the three vertices are charged with a unit of communication load. This stems from the broadcast nature of wireless communication and the need for local coordination in the trajectory based forwarding. The data found by a read access are delivered to the node that initiates the access, through a routing scheme (e.g., shortest path routing) independent of the quorum system. As a result, we do not consider the load introduced by data delivery in the simulation, as it is a constant for all quorum systems.

  \subsection{Comparing GeoQuorum with Existing Designs} \label{sec:comp}
    We compare GeoQuorum with $\mathcal{Q}_G$ and $\mathcal{Q}_L$ introduced in \cite{SarkarZG-MobiCom06}, as well as $\mathcal{Q}^m_G$, our remedy to $\mathcal{Q}_G$. Actually, a direct comparison is not possible, because though we faithfully take the quorum systems from \cite{SarkarZG-MobiCom06}, the underlying geometric design tools are different, as we explained in Sec.~\ref{sec:qsps} and \ref{sec:bkgccg}. Therefore, the following comparisons are based on our CCG design space. We assume WSNs with 5000 nodes. There is one data type, 500 nodes are contributing to it and 100 nodes may query it. We normalize the data query rate to 1 and vary the data production rate $r$ to test the system performance. Note that the actual write and read access rates (to a quorum system) are $500r$ and 100, respectively. Such an asymmetry between data production and consumption is reasonable, as otherwise multiple convergecasts may lead to better performance. For GeoQuorum, we set $R_W = 0.2\pi$ and $a = 0.2$. For each value of $r$, we obtain simulation results for 10 WSNs and we show the mean value and the standard deviation.

\begin{figure}[htb]
        \begin{center}
        \includegraphics[width=.85\columnwidth]{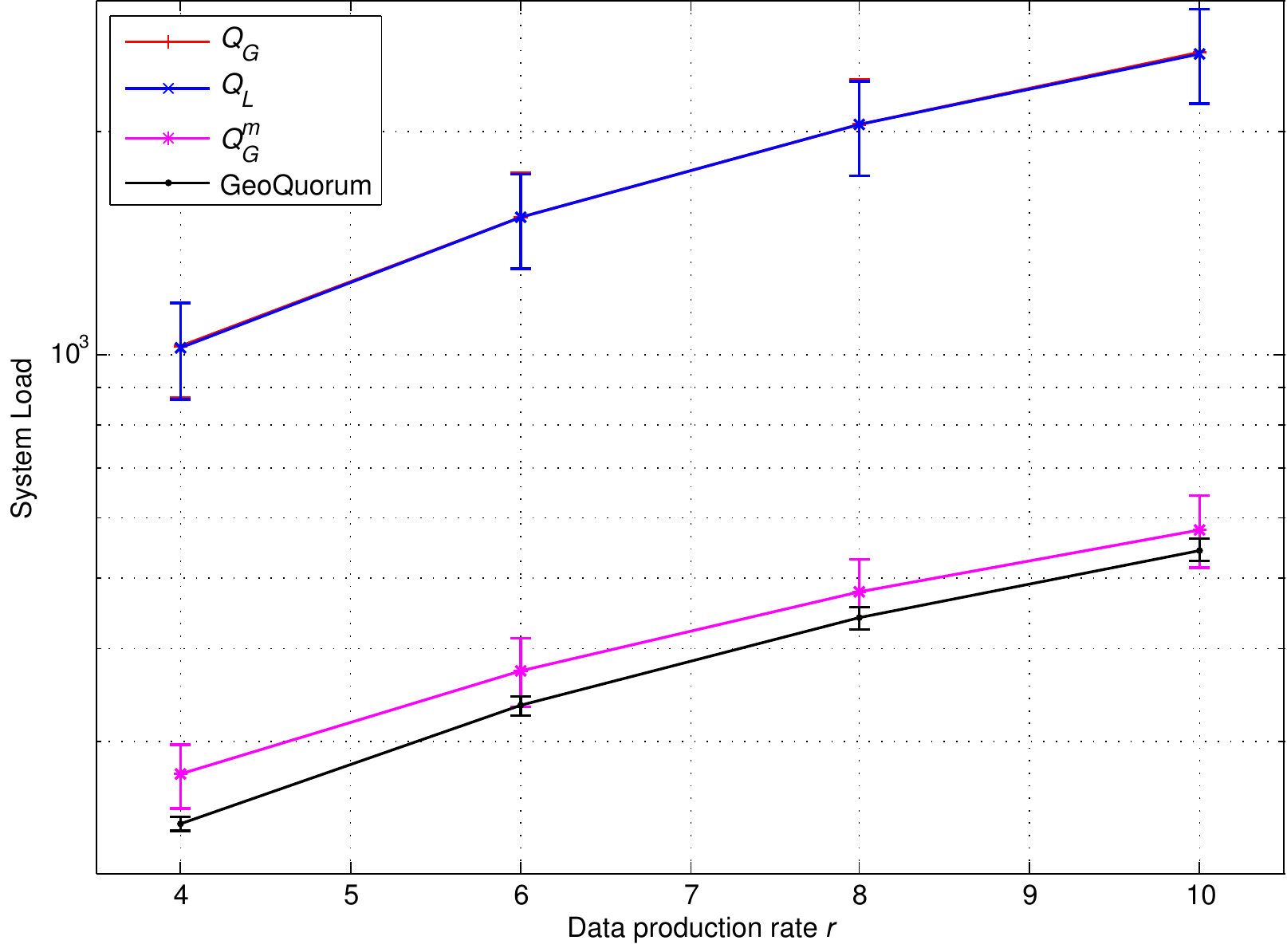}
        \caption{System load comparisons.}
        \label{fig:maxldcomp}
        \end{center}
        \vspace{-2ex}
    \end{figure}

\begin{figure}[htb]
        \begin{center}
        \includegraphics[width=.85\columnwidth]{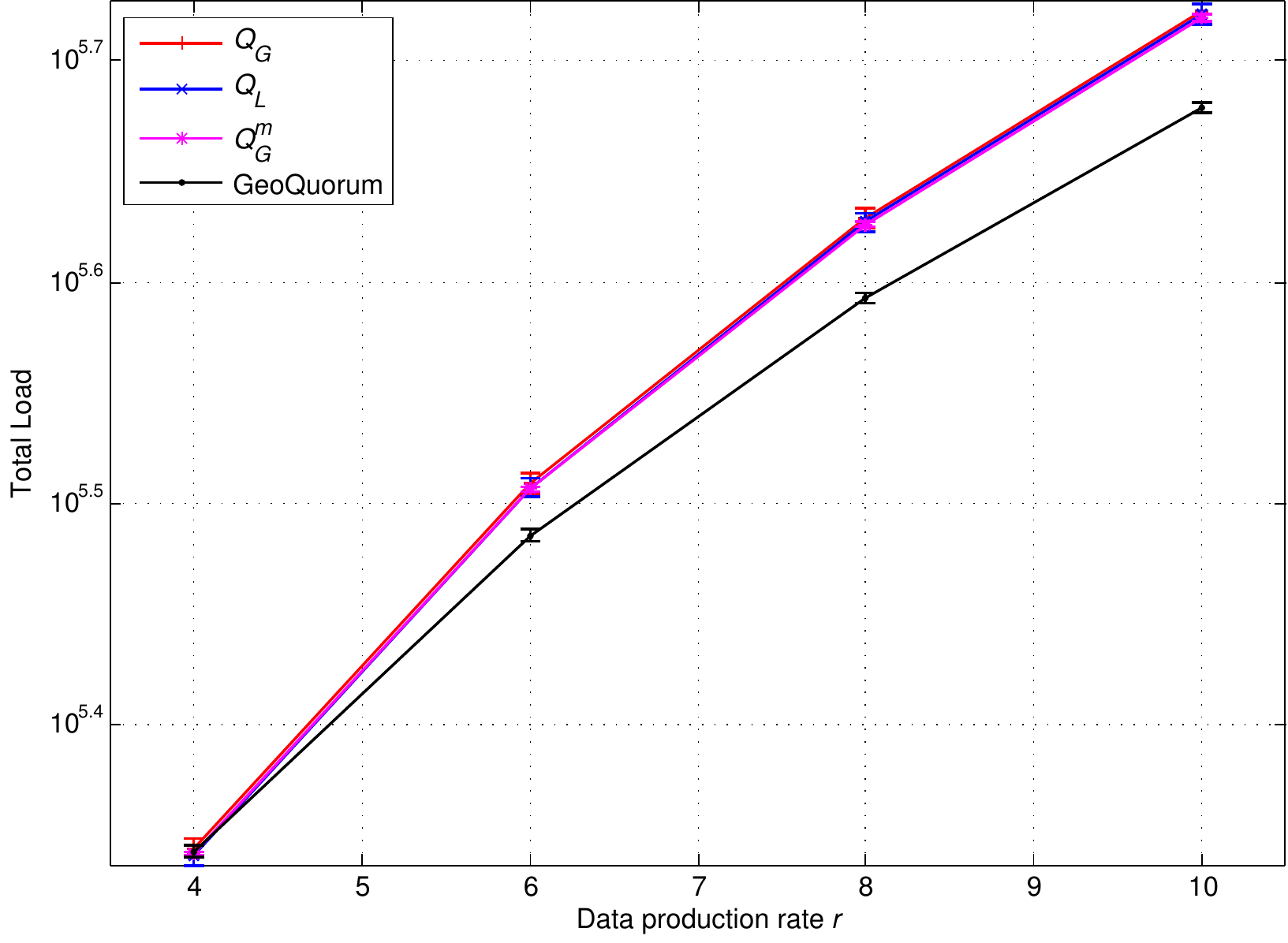}
        \caption{Total load comparisons.}
        \label{fig:ttldcomp}
        \end{center}
        \vspace{-2ex}
    \end{figure}

    We first compare the system load of the four quorum systems in Fig.~\ref{fig:maxldcomp}, then their total load in Fig.~\ref{fig:ttldcomp}.
    \begin{figure*}[htb]
    \parbox{\textwidth}{\parbox{.06\textwidth}{\center\includegraphics[width=.06\textwidth]{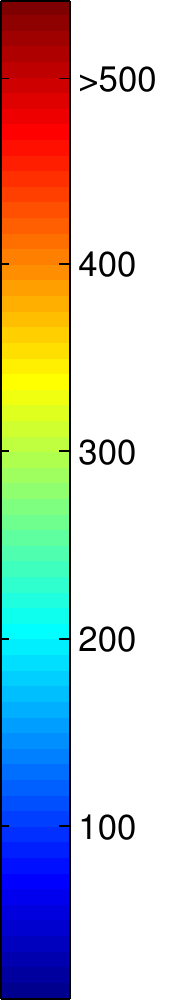}}
                        \parbox{.31\textwidth}{\center\includegraphics[width=.3\textwidth]{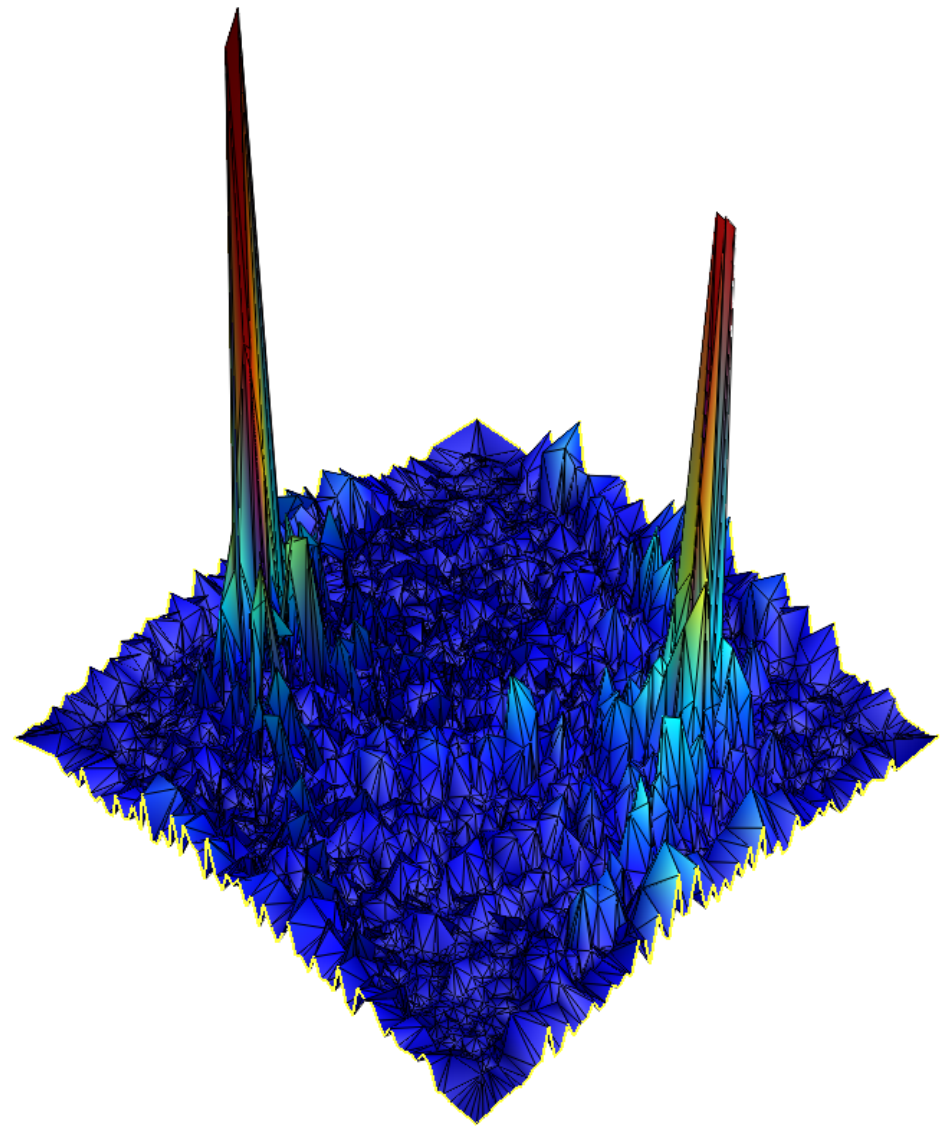}}
                        \parbox{.31\textwidth}{\center\includegraphics[width=.3\textwidth]{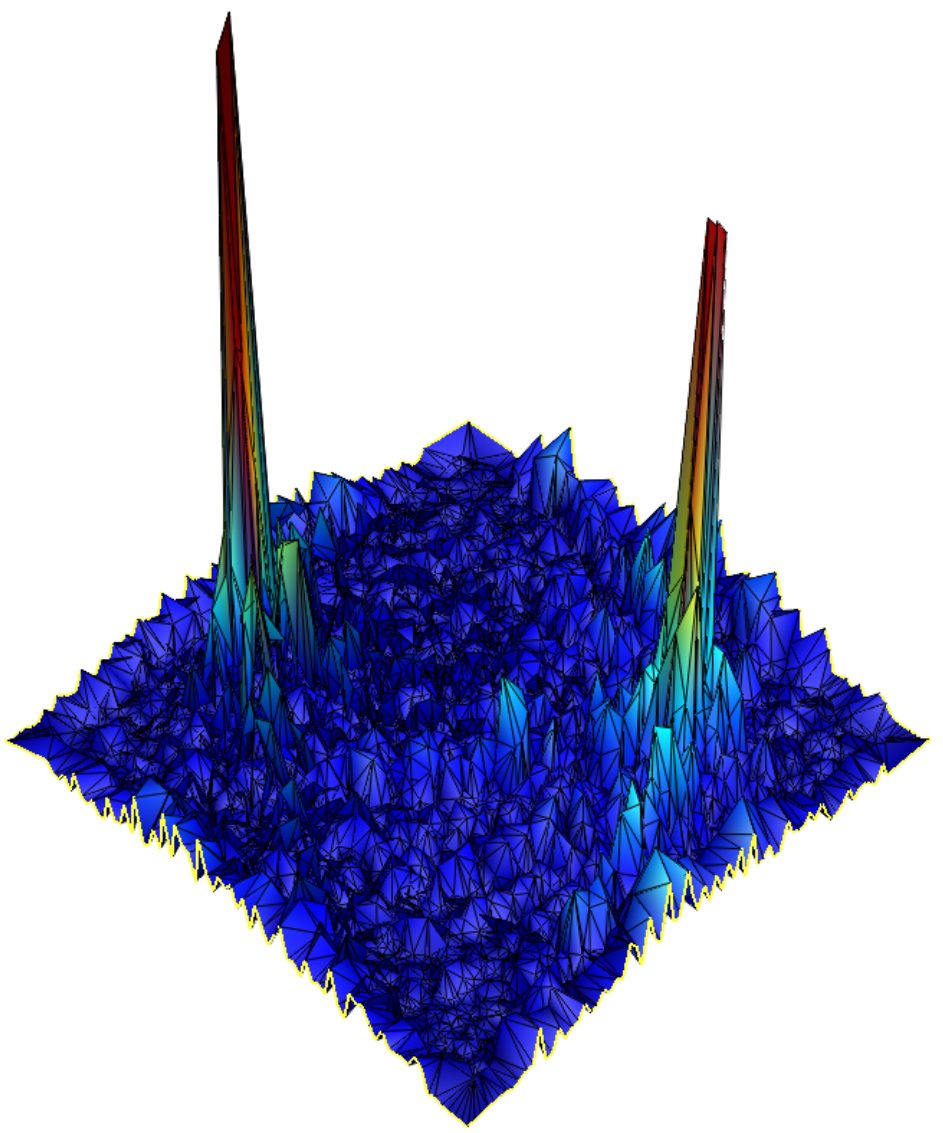}}
                        \parbox{.31\textwidth}{\center\includegraphics[width=.3\textwidth]{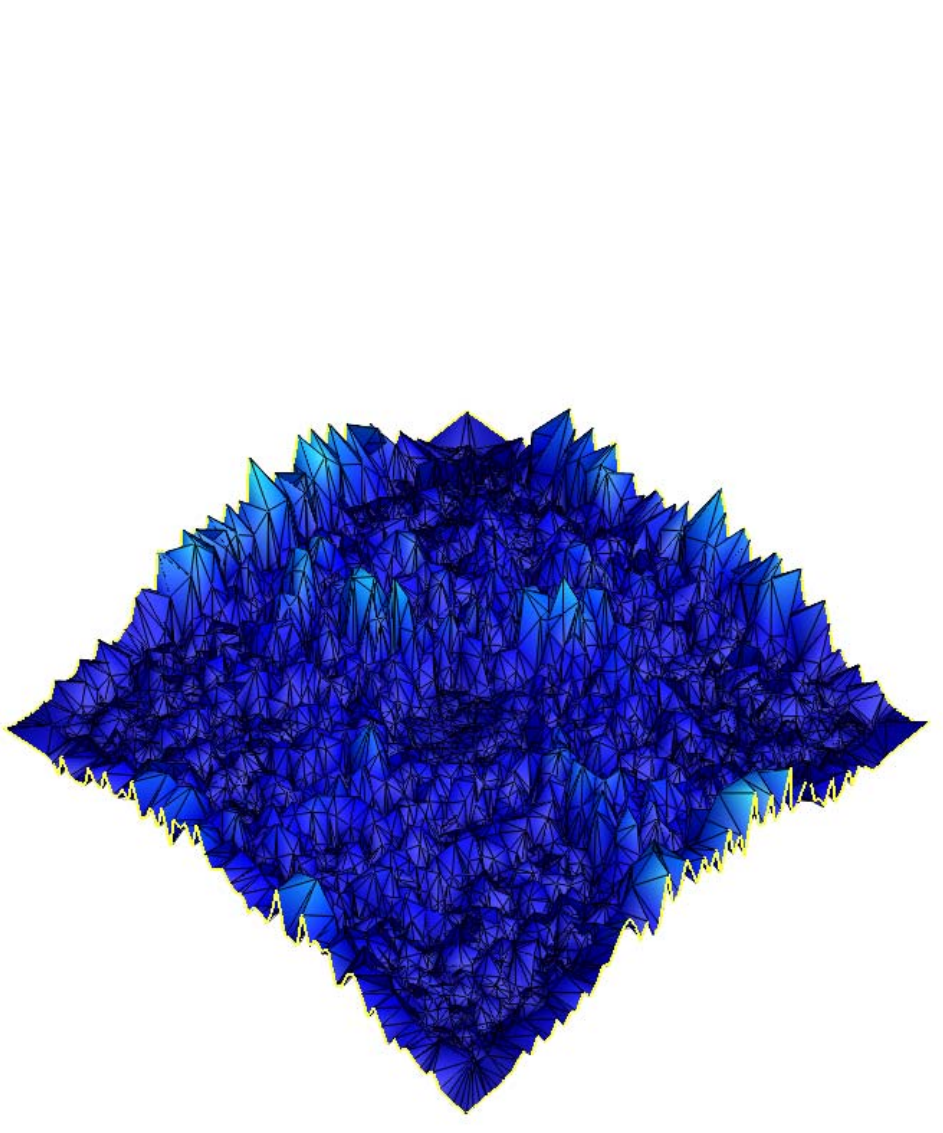}}}
    \parbox{\textwidth}{\parbox{.06\textwidth}{\ }
                        \parbox{.31\textwidth}{\center\scriptsize(a) $\mathcal{Q}_G$}
                        \parbox{.31\textwidth}{\center\scriptsize(b) $\mathcal{Q}_L$}
                        \parbox{.31\textwidth}{\center\scriptsize(c) GeoQuorum}}
    \caption{Illustration of load distribution for three different quorum systems.}
    \label{fig:distrib}
    \vspace{-1ex}
    \end{figure*}
    We illustrate the actual load distribution in Fig.~\ref{fig:distrib}. The following observations are immediate from these figures.
    \begin{itemize}
      \item The load distribution of $\mathcal{Q}_G$ and $\mathcal{Q}_L$ are very unbalanced, exactly due to the existence of a hash location $h$ and its antipodal point $\bar{h}$, as pointed out by \textit{Proposition~\ref{prp:load}}.
      \item Both GeoQuorum and $\mathcal{Q}^m_G$ significantly reduce the system load, as they benefit from using mixed access strategy.
      \item GeoQuorum incurs a much lower total load compared with all other three systems, due to its adaptivity to the asymmetry in data production and consumption.
    \end{itemize}
    Although the performance of $\mathcal{Q}_G$ and $\mathcal{Q}_L$ appear to be very similar in the figures, they actually differ by about $1\%$ to $2\%$. As explained in Sec.~\ref{sec:gqsals}, $\mathcal{Q}_L$ differs from $\mathcal{Q}_G$ only in read quorums, but as we assume that read access rate is far lower than the write access rate, this difference is ``diluted''.

  \subsection{Tuning the Load of GeoQuorum}
\begin{figure}[htb]
        \begin{center}
        \includegraphics[width=.85\columnwidth]{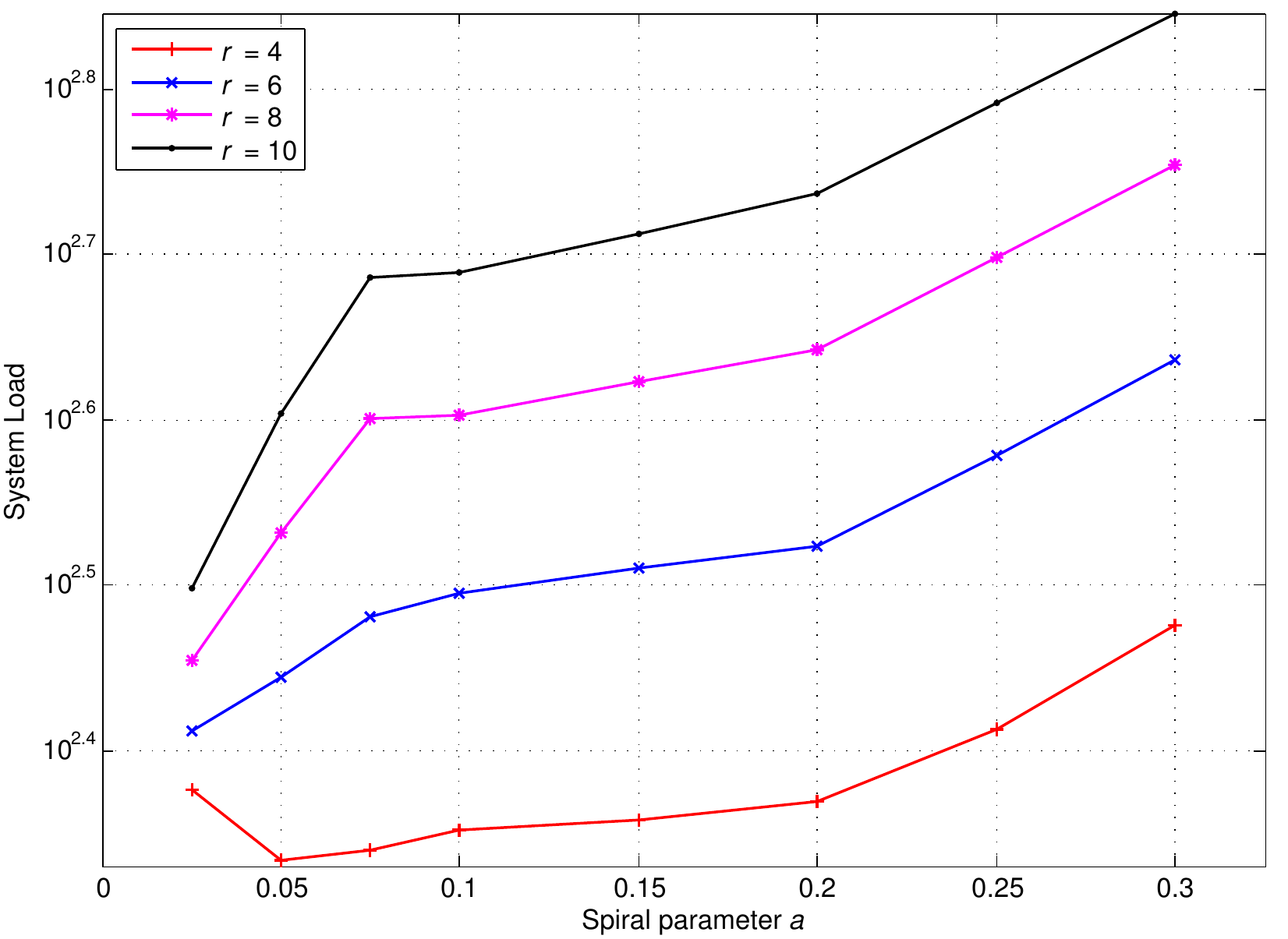}
        \caption{Tuning the system load.}
        \label{fig:maxldtn}
        \end{center}
        \vspace{-2ex}
    \end{figure}

\begin{figure}[htb]
        \begin{center}
        \includegraphics[width=.85\columnwidth]{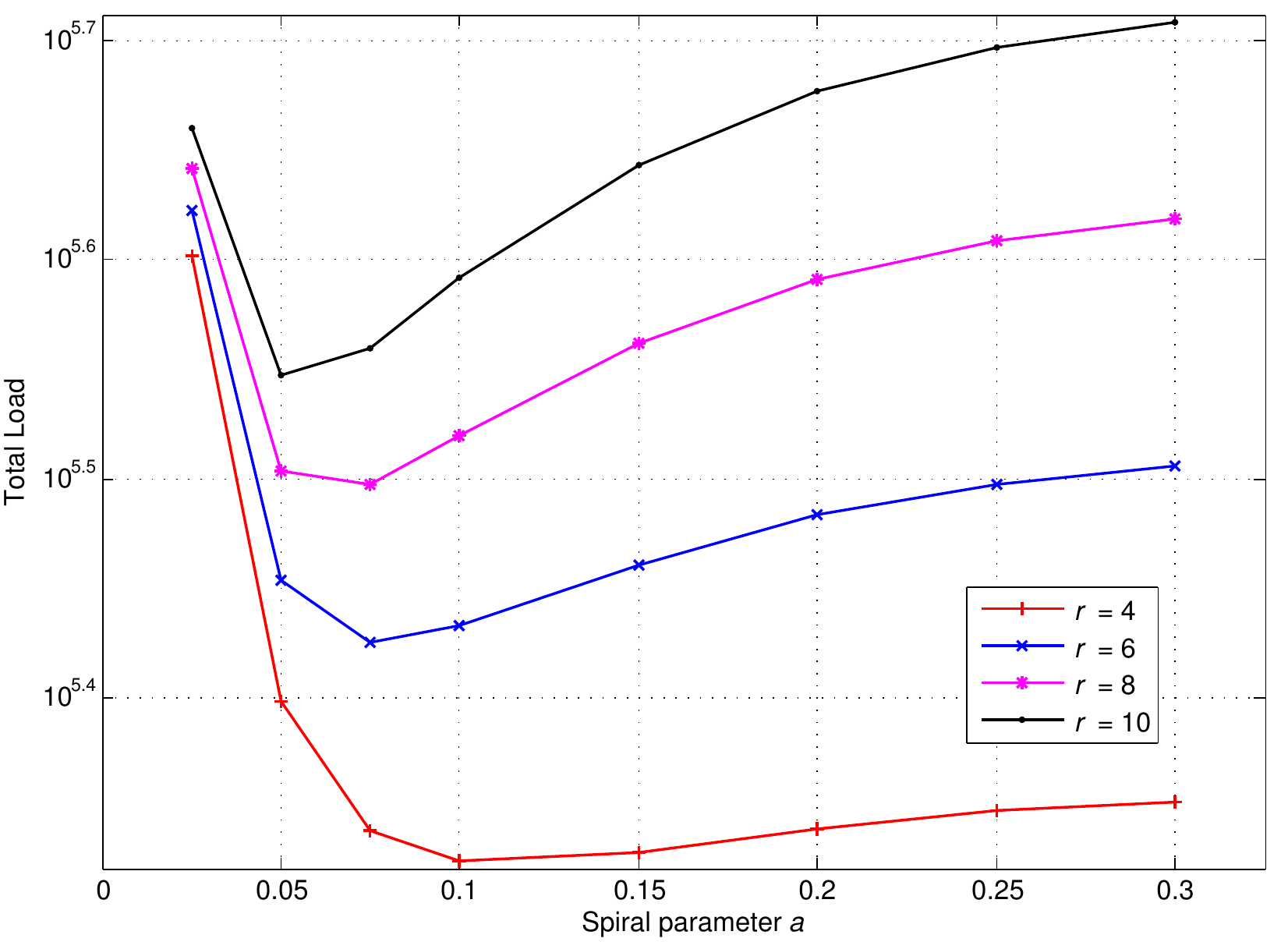}
        \caption{Tuning the total load.}
        \label{fig:ttldtn}
        \end{center}
        \vspace{-2ex}
    \end{figure}

    We show the performance of our GeoQuorum under parameter fine-tuning in this section. In particular, we use the same 5000-node WSNs in Sec.~\ref{sec:comp} and the same four values of $r$. We tune the spiral parameter $a$ from 0.025 to 0.3 while increasing $R_W$ proportionally to maintain the same robustness. The results on system and total load are plotted in Fig.~\ref{fig:maxldtn} and \ref{fig:ttldtn}, respectively. We only show mean values, as the standard deviations are very small (partially due to the load balancing effect brought by GeoQuorum).

    In general, one always has to make a tradeoff between load balancing and energy efficiency. The tunability of GeoQuorum allows us to make different tradeoffs upon different application requirements. For example, when the data production rate is low ($r = 4$), $a \in (0.75, 1.5)$ appears to achieves a balanced performance in both system and total load. This region shifts towards smaller values with an increasing $r$. For $r = 10$, $a$ is better to be around 0.05. The flexibility of freely tuning the system performance is one of the major advantages of GeoQuorum over the existing designs.

  \subsection{Tuning the Robustness of GeoQuorum}
\begin{figure}[htb]
        \begin{center}
        \includegraphics[width=.85\columnwidth]{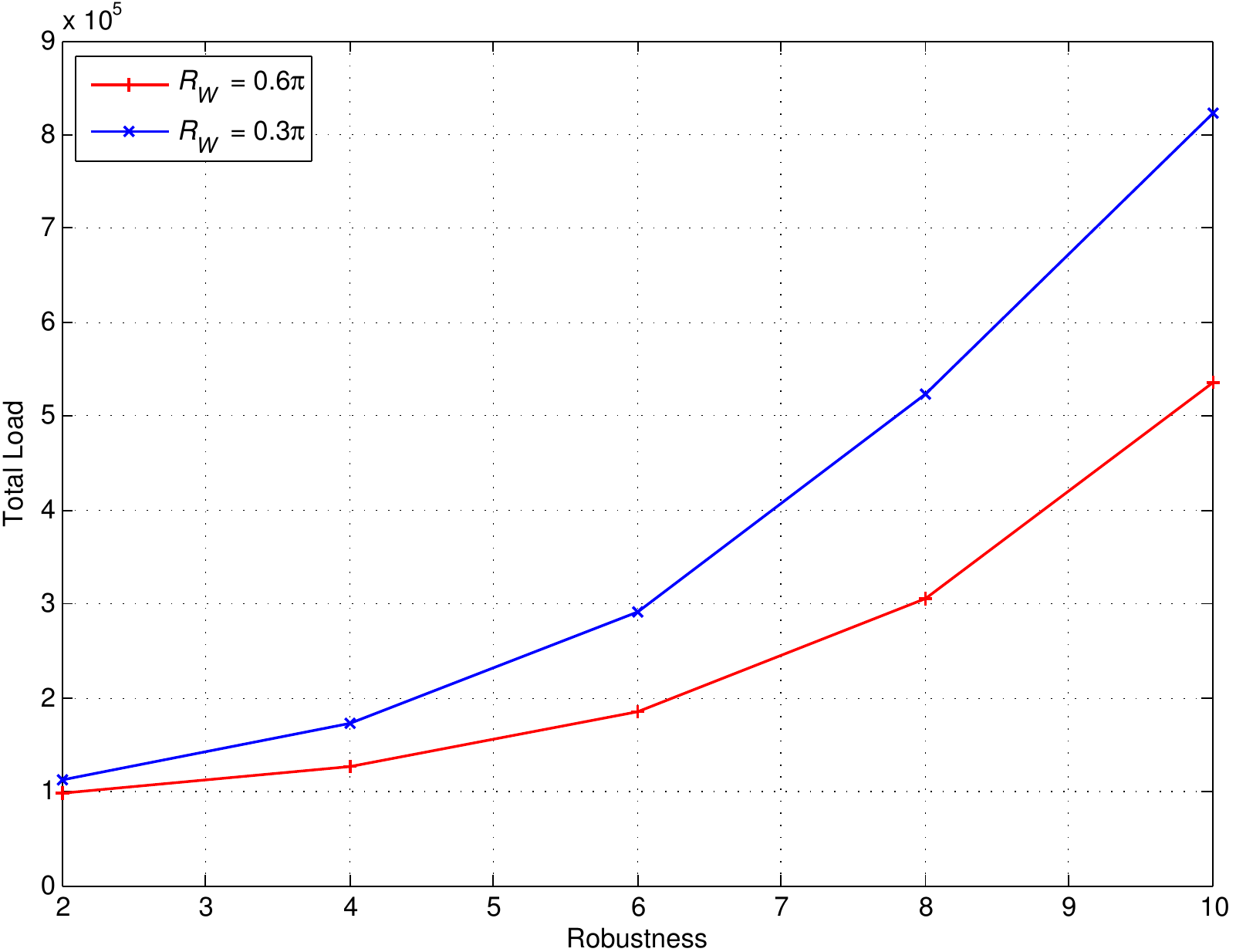}
        \caption{Tuning the robustness.}
        \label{fig:rbsttn}
        \end{center}
    \end{figure}

    If we just tune $a$ but keep $R_W$ constant, we will change the robustness of GeoQuorum. As shown by \textit{Proposition~\ref{prp:rbst}}, the robustness can be tuned at a granularity of 2. Of course, increasing robustness comes at a cost of an increased total load. Again using the 5000-node WSNs, we show the relation between robustness and total load by Fig.~\ref{fig:rbsttn}.
    We consider two cases where $R_W = 0.6\pi$ and $R_W = 0.3\pi$. When we tune $a$ to linearly increase the robustness from 2 to 10, the total load increases by (roughly) following a power law.

  \subsection{GeoQuorum in Irregular Network Area}
\begin{figure}[htb]
    \parbox{\columnwidth}{\parbox{.49\columnwidth}{\center\includegraphics[width=.45\columnwidth]{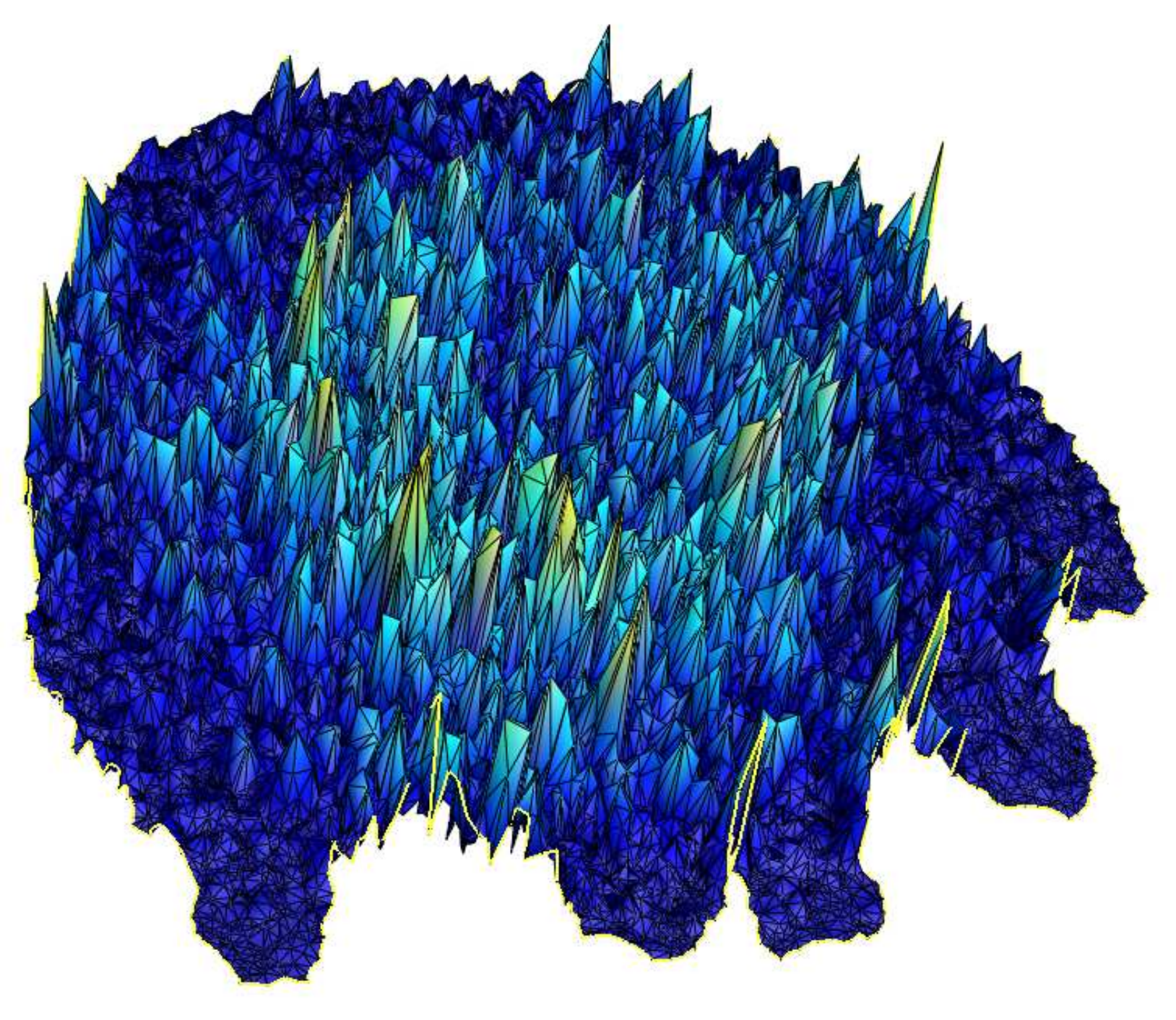}}
                          \parbox{.49\columnwidth}{\center\includegraphics[width=.45\columnwidth]{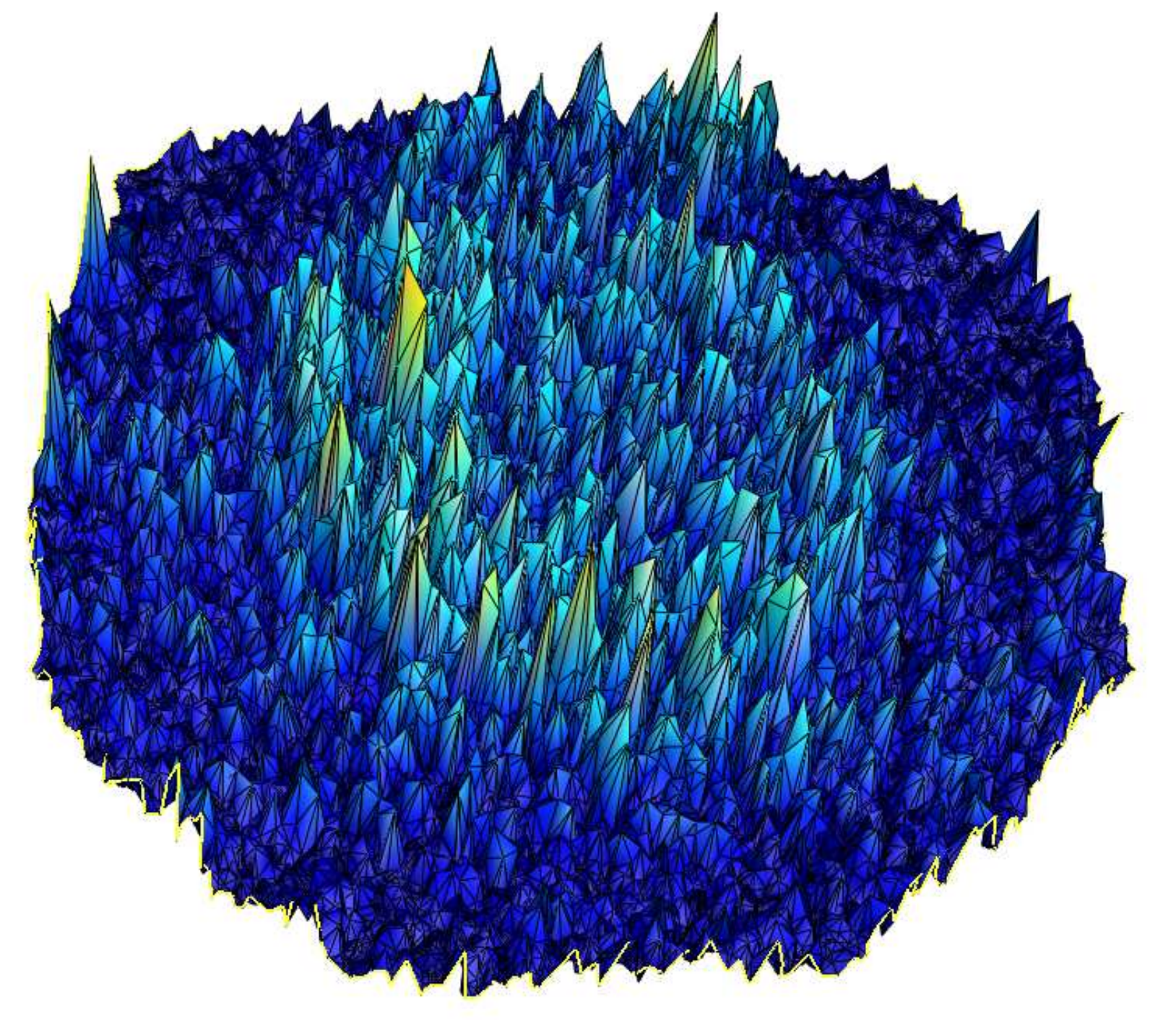}}}
    \caption{Load distribution of GeoQuorum on irregular areas.}
    \label{fig:idistrib}
    \end{figure}

    As all the previous simulations are based on 5000-node WSNs on square area, we simply demonstrate the ability of our design tools to cope with irregular areas in this section. Fig.~\ref{fig:idistrib} shows two WSNs deployed on irregular areas, with each consisting of 20000 nodes, one data type, 2000 contributors, 500 queriers.
    The load balancing effect of GeoQuorum is evident, though the load close to the network center is slightly higher than that close to the boundary. This slight unbalance is the cost one has to pay to maintain an acceptable total load: if one aims at fully balancing the load, larger quorums (longer curves) that make detour close to the boundary have to be used, leading to an unnecessarily high total load.

\section{Conclusion} \label{sec:con}
  We have investigated the issue of data access in WSNs, aiming at balancing (communication) load distribution while maintaining energy efficiency. Specifically, we have revived the application of quorum systems in WSNs, and proposed the concept of geometric quorum systems based on a new development in combining computational conformal geometry with sensor networking. In particular, we have proposed GeoQuorum that makes use of parameterized spatial curves to form quorums, such that the system performance can be fine-tuned to meet different application requirements. Through both analysis and simulations, we have confirmed the advantages of GeoQuorum over existing proposals.

  Our future work will focus on the implementation aspect of geometric quorum systems in general. We are planning to deploy these curve formed quorum systems in a WSN testbed, in order to obtain better insights on the performance and practicality of such systems.

\bibliographystyle{IEEEtran}
\bibliography{GeoQuorum}

\end{document}